\documentclass[11pt]{article}

\usepackage[letterpaper, left=1in, top=1in, right=1in, bottom=1in,nohead,includefoot]{geometry}
\usepackage{color,charter,graphicx,natbib,enumitem,amsmath,amssymb,amsfonts,titlesec,placeins,bm,pdfpages, subfig, placeins,setspace}
\usepackage[title]{appendix}
\usepackage{sidecap} \sidecaptionvpos{figure}{c} 
\usepackage[svgnames,x11names]{xcolor}
\usepackage[pdftex]{hyperref}
	\hypersetup{
	colorlinks,%
	linkcolor=MidnightBlue,  
	urlcolor=DarkSlateGray,   
	citecolor=RoyalBlue2  
	}
\titleformat*{\section}{\normalfont\Large\bfseries\blu}
\titleformat*{\subsection}{\normalfont\large\bfseries\blu}
\titleformat*{\subsubsection}{\normalfont\normalsize\bfseries\blu}
\def\para#1{\medskip\noindent{\bf #1}}
 
\def\blu{\color{RoyalBlue4}}      
\def\bzero{\mathbf{0}}
\def\A{\mathbf{A}}\def\B{\mathbf{B}}\def\C{\mathbf{C}}\def\D{\mathbf{D}}\def\G{\mathbf{G}}
\def\I{\mathbf{I}}\def\M{\mathbf{M}}\def\F{\mathbf{F}}\def\R{\mathbf{R}}\def\S{\mathbf{S}}
\def\W{\mathbf{W}}\def\H{\mathbf{H}}
\def\U{\mathbf{U}}\def\Z{\mathbf{Z}}
\def\e{\textrm{e}}\def\f{\mathbf{f}}\def\r{\mathbf{r}}
\def\y{\mathbf{y}}\def\z{\mathbf{z}}

\def\bSigma{{\bm\Sigma}}\def\bGamma{{\bm\Gamma}}
\def\bOmega{{\bm\Omega}}\def\bTheta{{\bm\Theta}}\def\bPsi{{\bm\Psi}} \def\bnu{{\bm\nu}}

\def\cD{\mathcal{D}} 
\def\seq#1#2{#1{:}#2}\def\j1J{j=\seq 1J}
\def\eqn#1{eqn.~(\ref{eq:#1})} \def\eqntwo#1#2{eqns.~(\ref{eq:#1},\ref{eq:#2})}

\def\etr{\textrm{etr}}

\def\bi{\begin{itemize}[itemsep=1pt,topsep=3pt]}
\def\ei{\end{itemize}}\def\i{\item}
\def\bna{\begin{enumerate}[itemsep=1pt,topsep=3pt,label=(\alph*)]}\def\en{\end{enumerate}}
\def\bn{\begin{enumerate}[itemsep=1pt,topsep=3pt]}\def\en{\end{enumerate}}
\def\beq#1{\begin{equation}\label{eq:#1}}\def\eeq{\end{equation}}
\def\beas{\begin{align*}}\def\eeas{\end{align*}}
\def\bea{\begin{align}}\def\eea{\end{align}}
 
\def\tN{\textrm{N}}\def\tMN{\textrm{N}}\def\tNIW{\textrm{NIW}}\def\tCNIW{\textrm{CNIW}}\def\tW{\textrm{W}}\def\tIW{\textrm{IW}}
\def\tE{\textrm{E}}\def\tMB{\textrm{MB}}
\def\cniwdf{s}  

\newcommand{\blind}0 

\begin{document}
 	 
\title{\blu Compositional dynamic modelling for causal prediction\\ in multivariate time series}  
\if0\blind
{ 
\author{Kevin Li\thanks{\href{mailto:kevin.li566@duke.edu}{kevin.li566@duke.edu}. 
                                     Li is a Statistical Science PhD candidate at Duke University. 
	   \newline \indent\,  
	   $^{\dagger}$\href{mailto:gtierney2@gmail.com}{gtierney2@gmail.com}. Research developed while 
	    Tierney was a Statistical Science PhD student at Duke University. 
	   \newline \indent\,  
           $^{\ddagger}$\href{mailto:ch.hellmayr@gmail.com}{ch.hellmayr@gmail.com}. 
             Research developed while Hellmayr was  at 84.51$^\circ$, Cincinnati, OH, U.S.A. 
	   \newline \indent\,  
	   $^{\#}$\href{mailto:mike.west@duke.edu}{mike.west@duke.edu}. 
	    Arts  \& Sciences Distinguished Professor  of Statistics \& Decision Sciences at Duke University. 
	   \newline \\
	    	Department of Statistical Science, Duke University, Durham, NC 27708-0251, U.S.A. 	 
	   }, 
	   \, Graham Tierney$^{\dagger}$, Christoph Hellmayr$^{\ddagger}$ \& Mike West$^{\#}$
          } 
} \fi  

\maketitle\thispagestyle{empty}\setcounter{page}0
   
\begin{center}{\blu Abstract}\end{center} 
Theoretical developments in sequential Bayesian analysis of multivariate dynamic models underlie new methodology for causal prediction.  This extends  the utility of existing  models with computationally efficient methodology, enabling routine exploration of Bayesian counterfactual analyses with multiple selected time series as synthetic controls.   Methodological contributions also define the concept of outcome adaptive modelling to monitor and inferentially respond to changes in experimental time series following   interventions designed to explore causal effects.   The benefits of  sequential analyses with time-varying parameter models for causal investigations are inherited in this broader  setting.  A case study in commercial causal analysis-- involving retail revenue outcomes related to marketing interventions-- highlights the methodological advances. 

\bigskip
\noindent{\it\blu Keywords:} 
Bayesian forecasting, Business revenue forecasting, Counterfactual prediction, Decouple/recouple,  Outcome adaptive models, Synthetic controls

\newpage
 
\section{Introduction\label{sec:intro}}
	
Developments in time series for causal analysis increasingly emphasize multivariate outcomes and the importance of flexible models for relationships among series. We focus on the use of synthetic controls and counterfactual forecasting, where a set of time series are jointly modelled prior to an intervention and
then partitioned into post-intervention \lq\lq experimental'' and \lq\lq control'' sets.  Series identified as controls are assumedly not impacted by the intervention; they thus  provide some opportunity to explain post-intervention experimental outcomes by control-experimental relationships unrelated to the intervention.  Other aspects of the post-intervention development of experimental series may then partly reflect causal responses. Predicting hypothetical counterfactual  outcomes assuming no causal effects provides the basis for assessing this. 

Bayesian methods for time series using the synthetic control approach initiated in \citet{brodersen2015inferring} and has been expanded in several directions recently~\citep[e.g.][]{menchetti2022estimating,papadogeorgou2023evaluating,antonelli2023heterogeneous}.  Additional related developments have brought dynamic latent factor models into the causal domain~\citep[e.g.][]{pang2022bayesian} though with, of course, major model specification and computational burdens.  Some of the challenges to these recent approaches that define our perspective relate to the needs to admit time-varying parameters, especially in patterns of dependencies over time within and between control and experimental series.  A related concern is the appropriate evaluation of full uncertainties about counterfactual outcomes that a fully Bayesian analysis can provide.  It is also important to focus on-- and enable-- sequential analysis to highlight the inherently important interests in monitoring for potential causal effects as new data arises. Further, the  reliance of some previous methods on computationally intensive methods such as MCMC is a limiting factor for both model development and in terms of scalability to larger numbers of series.  The recent introduction to the causal forecasting area of the traditional class of multivariate dynamic linear models (MVDLMs:~\citealp[][chap.~10]{PradoFerreiraWest2021}) in \cite{TierneyEtAl2024} addressed these desiderata in a restricted setting.  The restrictions require that synthetic control series be defined as exogenous, used as predictors of both control and counterfactual series pre- and post-intervention.    This is very limiting and obviates a main goal here: that of using the full subset of series that are not subject to the intervention as the controls themselves.  A model framework that allows this will provide flexibility in understanding-- and utilizing-- the predictive value of the identified control series without imposing assumptions as to how they relate to the experimental series. This will, of course, require that pre-intervention dependencies  
are appropriately characterized.  

We address these goals by extending the theory and resulting methodology of the MVDLM framework. A simulated example conveys core ideas, and summaries from a case study in a business application related to retail marketing interventions highlights the methodological advances.

\section{Multivariate Dynamic Models\label{sec:MVDLM}}
\subsection{MVDLM Structure\label{sec:vanilla_mvdlm}} 
The column $q-$vector time series $\y_t$  follows a {\em multivariate dynamic linear model} (MVDLM) from the standard class of such models~\citep[][chap.~10]{PradoFerreiraWest2021}. At any time $t$
denote by $\cD_t$ all past data and information up to, and including, time $t$.   The MVDLM at time $t$ conditional on $\cD_{t-1}$ has the state-space form
\bea \label{eq:mvdlm}
\begin{aligned} 
\y_t' &= \F_t' \bTheta_t + \bnu_t',  \qquad  & \bnu_t &\sim \tN(\bzero, \bSigma_t), \\
\bTheta_t &= \G_t \bTheta_{t-1} + \bOmega_t,  \qquad & \bOmega_t  &\sim \tN(\bzero, \W_t, \bSigma_t), 
\end{aligned}
\end{align}
with the following elements:
\bi \itemsep-3pt
\i $\F_t,$ the $p-$vector of constants and potentially time-varying, exogenous predictors. 
\i  $\bTheta_t, $ the $p \times q$ state parameter matrix.
\i  $\bnu_t$, the observation noise $p-$vector with time-varying variance matrix $\bSigma_t$.
\i  $\G_t$, the  $p\times p$ state matrix  governing the structural part of the evolution of  $\bTheta_t$. 
\i  $\bOmega_t$, the matrix normal state innovation noise;  time-varying dependence influences stochastic changes over time in state parameters within each series ($p\times p$ variance matrix $\W_t)$ and across series ($q\times q$ variance matrix $\bSigma_t).$
\i $\W_t$ is parsimoniously defined by one or more specified state discount factors; our application uses a single discount factor $\delta\in (0,1].$ 
\i $\bSigma_t$ follows the Markovian volatility process, $\bSigma_t = \tMB_t(\bSigma_{t-1}),$ conditional on $\cD_{t-1},$ the standard
 matrix beta stochastic model for time-varying variance matrices~(Appendix~\ref{app:MatrixBeta}), 
defined by a {\em volatility discount factor} $\beta\in (0,1].$ 
\ei 
Univariate series in $\y_t$ follow dynamic linear models (DLMs) with the same $\F_t,\G_t,\W_t$ across series; the state vector for univariate series $j$ is the $j^{th}$ column of $\bTheta_t.$  Choices of $\F_t,\G_t$ model structural behaviour such as trends, seasonality and dynamic regressions, including vector autoregressions (VAR) and time-varying VAR models of particular relevance in economic applications~\citep[e.g.][sect.~9.4]{BanburaEtAl2010,KoopKorobilis2010,KoopKorobilis2013,PradoFerreiraWest2021}.
Examples here use constant $\F,\G$ to model local trends; relationships in trends across  series arise through $\bSigma_t$.   
 
\subsection{Bayesian Analysis in MVDLMs\label{sec:Bayes_mvdlm}}  
These models define fully conjugate forward filtering and forecasting analysis, enabling fast analytical computations and flexibility in adapting to time variation in cross-series dependencies.  The sequential filtering analysis--  time $t-1$ to $t$ evolution followed by Bayesian updating on observing $\y_t$--   involve (matrix) normal, inverse Wishart distributions (NIW; Appendix~\ref{app:allCNIW}).  Assuming a conjugate NIW distribution  $p(\bTheta_0,\bSigma_0|\cD_0)$,  the following summaries of four key steps arise for all $t\ge 1.$ 
 
\para{1. Posterior at time $t-1$:}   This is $(\bTheta_{t-1}, \bSigma_{t-1} | \cD_{t-1}) \sim \tNIW(\M_{t-1}, \C_{t-1}, n_{t-1}, \D_{t-1})$ 
with components
$(\bTheta_{t-1}|\bSigma_{t-1},\cD_{t-1}) \sim \tMN(\M_{t-1}, \C_{t-1}, \bSigma_{t-1})$ and $\bSigma_{t-1}|\cD_{t-1} \sim \tIW(n_{t-1},\D_t).$

\para{2. Prior for time $t$:}   Evolving to time $t$, the implied prior for time $t$ parameters is
 $(\bTheta_t, \bSigma_t |\cD_{t-1}) \sim \tNIW( \M_t^*, \C_t^*, n_t^*, \D_t^*)$ with 
 $\M_t^* = \G_t \M_{t-1} $, $\C_t^* = \G_t \C_{t-1} \G_t'+\W_t$, $n_t^*=\beta n_{t-1}-(1-\beta)(q-1)$ and $\D_t^*=\beta\D_{t-1}.$  
Using a single state discount factor $\delta\in (0,1]$ implies that $\C_t^* = \G_t \C_{t-1} \G_t'/\delta$.  
Values of $\delta, \beta$ closer to one reflect more stability over time in $\bTheta_t, \bSigma_t$ respectively. 

\para{3. Forecasting for time $t$:}  $p(\y_t|\cD_{t-1})$ is multivariate T with d.o.f. $n_t^*$, 
location vector $\f_t$ where $\f_t' =  \F_t'\M_t^*$, 
and scale matrix $q_t\S_t$ where $q_t = 1 + \F_t'\C_t^*\F_t$ and $\S_t = \D_t^*/n_t^*.$  This is easily simulated 
based on sampling $p(\bTheta_t, \bSigma_t |\cD_{t-1}).$ Forecasting more than $1-$step ahead recurses this into the future.  

\para{4. Posterior at time $t$:} Observing $\y_t$ updates information to $\cD_t = \{ \y_t, \cD_{t-1}\}$ with the implied conjugate posterior
$(\bTheta_t, \bSigma_t |\cD_t) \sim \tNIW(\M_t, \C_t, n_t, \D_t)$. With $q-$vector point forecast error 
$\e_t = \y_t - \f_t$ and $q-$vector of adaptive coefficients $\A_t = \C_t^*\F_t/q_t,$ the 
 updated parameters are 
$\M_t = \M_t^* + \A_t\e_t'$, $\C_t = \C_t^* -\A_t\A_t'q_t$, $n_t = n_t^*+1$ and $\D_t = \D_t^*+\e_t\e_t'/q_t.$ 

\newpage

\section{Compositional Dynamic Modelling for Causal Prediction\label{sec:causaldynamicmodels}}

\subsection{Counterfactual Setting, Synthetic Controls and Analysis Goals\label{sec:causalsetting}} 
The setting involves an intervention time point $T$ when a subset of the series are identified as experimental, 
potentially impacted by the intervention. The assumption is that the remaining series are not impacted, and to be regarded as controls. With no loss of generality write $\y_t'=(\y_{ct}',\y_{et}')$ for the partition of the $q$ series into the vertical stack of the $q_c-$vector $\y_{ct}$ and the $q_e-$vector $\y_{et}$, where $q_c+q_e=q.$  The notation reflects the causal interests with $c$ denoting  series that will be designated as 
control and $e$ experimental when $t\ge T.$ We use the same notation and terminology for $t<T.$  In the common observational setting, the controls are regarded as synthetic, chosen on the assumption of no impact of the intervention, and no interference (spillover) of causal effects  post-intervention.   
Pre-intervention modelling analyses all series jointly, characterizing potentially time-varying 
cross-series dependencies. Post-intervention, analysis continues in the counterfactual setting where only $\y_{ct}$ is observed and predictions of hypothetical unobserved, counterfactual outcomes  are made to compare with the actual  outcomes. 

Specifically, write $\y_{e_0t}$ for latent, counterfactual outcomes for $t\ge T;$ this is the hypothetical development of the designated experimental series if the intervention has no effect.  Pre-intervention, both $\y_{ct}$ and $\y_{e_0t}=\y_{et}$ are observed, while post-intervention $\y_{ct}$ is observed but $\y_{e_0t}$ is \lq\lq missing data''.  Thus $\y_{e_0t}$ is regarded a latent series distinct from the actual post-intervention outcomes on the $e-$series, and no information from the latter is using in model analyses for $t\ge  T.$  

MVDLMs are workhorses of applied time series monitoring and forecasting. They have recently been applied in causal studies in settings where the $\F_t$ vector has exogenous predictors chosen as synthetic controls~\citep{TierneyEtAl2024,TierneyEtAl2024Supplement}. There, predictor variables  are chosen as elements of $\F_t$ to reflect and partly explain relationships across subsets of series in $\y_t$ based on data observed prior to the intervention time. Of those, predictors identified as controls are assumed to be unaffected by the intervention; hence, post-intervention, they enable partial  correction for cross-series dependencies that are putatively unrelated to the intervention. Then, post-intervention development of the series can be interrogated for additional patterns that may plausibly be identified as causal responses.    Important features of the MVDLM analysis-- and core to our goals here-- are that: 
\begin{itemize} \itemsep-3pt
\item  cross-series dependencies  and regression effects of synthetic controls can change over time;  
\item  causal effects may change over the post-intervention period,  requiring appropriate uncertainty characterization when predicting the  counterfactual series; 
\item in some applications, interests also concern causal implications for cumulative and/or aggregate effects on the experimental series, and a fully Bayesian multivariate model amenable to direct evaluation of posterior and predictive distributions is needed; 
\item  analysis is sequential and time-adaptive, so that post-intervention causal inferences can be monitored as control data accrues;
\item analysis is fully conjugate, theoretically precise and  computationally scalable.
\end{itemize}
A main focus of the current paper is to use the full $\y_{ct}$ as controls along with any other chosen, exogenous predictors in the $\F_t$ regression vectors.  However, the  MVDLM theory breaks down in this setting. That theory requires full observations on all series in $\y_t$. In the counterfactual framework, the subvector $\y_{e_0t}$ is unobserved for $t\ge T$ so that there is no post-intervention information relevant to inferences on relevant columns of the state matrix 
$\bTheta_t$ or for variances and covariances in $\bSigma_t$ related to the experimental series. At time $t=T,$ the posterior for $(\bTheta_T,\bSigma_T|\cD_T)$ is no longer of the NIW form, and that is then true for all future priors and posteriors as the model evolves.  Hence, generalization 
of the existing MVDLM analysis is needed.  

\subsection{Compositional Representation of Traditional MVDLM}
We first recapitulate the standard MVDLM analysis with the model, priors and posteriors represented in compositional forms that 
reflect the structure $\y_t'=(\y_{ct}',\y_{et}').$ Conformably partition $(\bTheta_t,\bSigma_t)$ in  \eqn{mvdlm} as
$$ 
\bTheta_t = (\bTheta_{ct},\bTheta_{et})\quad\textrm{and}\quad  \bSigma_t = \begin{pmatrix} \bSigma_{ct}&\bSigma_{ect}'\\ \bSigma_{ect}&\bSigma_{et}\end{pmatrix} $$  
with implicit dimensions. This identifies the first $q_c$ columns of $\bTheta_t$ as $\bTheta_{ct},$ the state vectors of control series, and 
the latter $q_e$ columns $\bTheta_{et}$ as those of experimental series, with corresponding partitioning of the cross-series variance 
matrix $\bSigma_t$.
The theory and notation of Appendix~\ref{app:allCNIW} now applies with the time $t$ index added. 

\paragraph{Observation equation:} 
The first component of the  MVDLM  in \eqn{mvdlm} defines the observation p.d.f. $p(\y_t|\bTheta_t,\bSigma_t,\cD_{t-1}).$ 
This can be re-expressed in compositional form with $c-$marginal and $e|c-$conditional distributions 
\begin{equation}\label{eq:compMVDLMyt}
\begin{split}
(\y_{ct}'|\bTheta_{ct},\bSigma_{ct},\cD_{t-1}) & \sim \tN(\F_t'\bTheta_{ct},\bSigma_{ct}),\\  
(\y_{et}'|\y_{ct},\bTheta_{ct},\bTheta_{et},\bGamma_{et},\bPsi_{et},\cD_{t-1}) &   \sim \tN( \F_t'\bTheta_{et} + (\y_{ct}'-\F_t'\bTheta_{ct})\bGamma_{et}', \bPsi_{et}),
\end{split}
\end{equation}
where
$\bGamma_{et} = \bSigma_{ect}\bSigma_{ct}^{-1}$ and $\bPsi_{et} =  \bSigma_{et} - \bSigma_{ect}\bSigma_{ct}^{-1}\bSigma_{ect}'.$ 

\paragraph{State evolution equation:} 
The second component of the  MVDLM  in \eqn{mvdlm} defines the state evolution p.d.f.  $p(\bTheta_t|\bTheta_{t-1},\bSigma_t,\cD_{t-1}).$ 
This can be re-expressed in compositional form with $c-$marginal and $e|c-$conditional distributions 
\begin{equation}\label{eq:compMVDLMThetat}
\begin{split}
(\bTheta_{ct}|\bTheta_{c,t-1},\bSigma_{ct},\cD_{t-1}) & \sim \tN(\G_t\bTheta_{c,t-1},\W_t,\bSigma_{ct}),\\  
(\bTheta_{et}|\bTheta_{ct},\bTheta_{e,t-1},\bGamma_{et},\bPsi_{et},\cD_{t-1})   
  &   \sim \tN( \G_t\bTheta_{e,t-1} + (\bTheta_{ct}-\G_t\bTheta_{c,t-1})\bGamma_{et}', \W_{et},\bPsi_{et}),
\end{split}
\end{equation}
where
$\bGamma_{et}, \bPsi_{et}$ are defined above, and $\W_{et}\equiv \W_t$ is the same within-column state variance matrix in the $e|c-$conditional evolution as in the $c-$marginal evolution. 

\paragraph{Volatility matrix evolution equation:} 
The process $\bSigma_t = \tMB(\bSigma_{t-1})$ defines independent Markov evolutions 
of the $c-$marginal and $e|c-$conditional parameters  $\bSigma_{ct}$ and $(\bGamma_{et},\bPsi_{et}).$   Partition $\D_t^*$ of the time $t$ prior $\bSigma_t \sim \tIW(n_t^*,\D_t^*)$
conformably with that of $\bSigma_t.$   The time $t$ prior 
$p(\bSigma_{ct},\bGamma_{et},\bPsi_{et}|\cD_{t-1})$ has compositional form with $c-$marginal and $e|c-$conditional distributions 
\begin{equation}\label{eq:compMVDLMSigma}
\begin{split}
\bSigma_{ct}|\cD_{t-1} & \sim \tIW(n_t^*,\D_{ct}^*)\ \textrm{arising from the marginal process} \ \bSigma_{ct}=\tMB(\bSigma_{c,t-1}),\\
\bGamma_{et}|\bPsi_{et}, \cD_{t-1}& \sim \tMN(\D_{ect}^*\D_{ct}^{*-1}, \bPsi_{et}, \D_{ct}^{*-1})\ \ \textrm{and} \  \ 
\bPsi_{et}| \cD_{t-1} \sim \tIW(n_t^*+q_c, \D_{et}^* - \D_{ect}^*\D_{ct}^{*-1}\D_{ect}^{*'}).
\end{split}
\end{equation}
 
\subsection{An Extended Class of Compositional MVDLMs\label{sec:compositionalMVDLMs}}

The compositional form of the MVDLM admits a more general class of conjugate priors.  This extended theory
is general but particularly suited to the context of counterfactual modelling for causal prediction-- a setting in which different levels of information arise for the $c-$marginal and $e|c-$conditional series where the standard theory breaks down. 
The development exploits the conditional decoupling of these two subseries. 
The compositional model  is that of \eqntwo{compMVDLMyt}{compMVDLMThetat}. One key point is that the evolution model of 
\eqn{compMVDLMThetat} allows the 
within-column state variance matrix $\W_{et}$ in the $e|c-$conditional evolution to differ from that $\W_t$ in the $c-$marginal evolution. This is important in extending the theory to allow for different covariance patterns-- and their changes over time-- for state vectors in the experimental series compared to those in the control series. Some of the essential underlying theory was exploited in simpler versions of MVDLMs by~\cite{Corradi1993} to address missing data on some of the series. The same concepts-- and our novel extensions of the theoretical results-- apply to the counterfactual setting since the $e$ series is effectively missing data post-intervention. 

The following four subsections correspond to, and generalize, the four items  $1{-}4$ in Section~\ref{sec:Bayes_mvdlm}. That standard MVDLM analysis is a special case of this more general compositional MVDLM framework.  Notation and terminology of CNIW and NIW-CNIW distributions follow Appendix~\ref{app:allCNIW} that  includes proofs of key theoretical results.  

\subsubsection{Conjugate Time $t-1$ Posterior}

Suppose that the time $t-1$ posterior is  the NIW-CNIW distribution with p.d.f. 
\beq{CNIWtimetminusoneposterior}  
      p(\bTheta_{c,t-1},\bSigma_{c,t-1}|\cD_{t-1}) 
         p(\bTheta_{e,t-1},\bGamma_{e,t-1}, \bPsi_{e,t-1}|\bTheta_{c,t-1},\cD_{t-1}) 
\eeq
having components
\begin{equation}\label{eq:comptimetminusoneposterior}
\begin{split}
\bTheta_{c,t-1},\bSigma_{c,t-1}|\cD_{t-1} &\sim \tNIW(\M_{c,t-1},\C_{t-1},n_{t-1},\D_{t-1}),\\
\bTheta_{e,t-1},\bGamma_{e,t-1}, \bPsi_{e,t-1}| \bTheta_{c,t-1}, \cD_{t-1}  &\sim  \tCNIW(\Z_{t-1}, \C_{e,t-1}, \cniwdf_{e,t-1}, \H_{t-1} | \bTheta_{c,t-1}).
\end{split}
\end{equation}
Defining parameters are known at time $t-1$. The $c-$marginal parameters are as in the usual MVDLM now applied in the $\y_{ct}$ model. The $e|c-$conditional has parameters as follows:
 \begin{itemize} \itemsep-3pt 
 \item a $p\times q$ matrix $\Z_{t-1} = (\Z_{c,t-1},\Z_{e,t-1})$ where $\Z_{c,t-1}$ is $p\times q_{c,t-1}$  and $\Z_{e,t-1}$ is $p\times q_{e,t-1}$;
 \item a $p\times p$ s.p.d. matrix $\C_{e,t-1}$;
 \item  a d.o.f. $\cniwdf_{e,t-1}>0$; 
 \item a $q\times q$ s.p.d. matrix $\H_{t-1}$ partitioned conformably with $\bSigma_{t-1}$ into
  s.p.d. diagonal blocks $\H_{c,t-1},\H_{e,t-1}$ and lower left off-diagonal block $\H_{ec,t-1}.$
\end{itemize}  

\subsubsection{Evolution to Conjugate Time $t$ Prior}
Independent evolutions of $(\bTheta_{et},\bSigma_{ct})$ and  $(\bTheta_{ct},\bGamma_{et}, \bPsi_{et})$ yield the time $t$ NIW-CNIW prior  
$$ 
      p(\bTheta_{ct},\bSigma_{ct}|\cD_{t-1}) 
         p(\bTheta_{et},\bGamma_{et}, \bPsi_{et} | \bTheta_{ct},\cD_{t-1})
$$
with components
\begin{equation}\label{eq:CNIWtimetprior}
\begin{split}
\bTheta_{ct},\bSigma_{ct}|\cD_{t-1} &\sim \tNIW(\M_{ct}^*,\C_t^*,n_t^*,\D_t^*),\\
\bTheta_{et},\bGamma_{et}, \bPsi_{et}| \bTheta_{ct}, \cD_{t-1}  &\sim  \tCNIW(\Z_t^*, \C_{et}^*, \cniwdf_{et}^*, \H_t^* | \bTheta_{ct}).
\end{split}
\end{equation}
The usual evolution equations apply in the $c-$margin:  $\M_t^* = \G_t \M_{t-1} $, $\C_t^* = \G_t \C_{t-1} \G_t'+\W_t$, $n_t^*=\beta n_{t-1}-(1-\beta)(q-1)$  and $\D_t^*=\beta\D_{t-1}$ using volatility discount factor $\beta;$ use of 
a single state discount factor $\delta\in (0,1]$ to define $\W_t$ implies that $\C_t^* = \G_t \C_{t-1} \G_t'/\delta$.   
The evolved quantities in the $e|c-$conditional have  analogous forms: $\Z_t^*=\G_t\Z_{t-1}$, $\C_{et}^*=\G_t\C_{e,t-1}\G_t'+\W_{et}$, 
$\cniwdf_{et}^* = \beta_e\cniwdf_{e,t-1}-(1-\beta_e)(q-1)$ and  $\H_t^*=\beta_e\H_{t-1}$ using volatility discount factor $\beta_e$; use of 
use of a single state discount factor $\delta_e\in (0,1]$ to define $\W_{et}$ implies that $\C_{et}^* = \G_t \C_{e,t-1} \G_t'/\delta_e$.   

The  decoupling allows different discount factors $(\beta_e,\delta_e)$ in the $e|c-$conditional model and $(\beta,\delta)$  in the $c-$margin.  This is of general interest and relevant in causal prediction as shown below.

\subsubsection{Forecasting at Time $t-1$} 
One-step ahead predictions under this compositional dynamic model are trivially quantified via direct simulation.  The modeller first draws direct Monte Carlo samples 
for all dynamic parameters from the prior in \eqn{CNIWtimetprior}. This involves simulation from normal and inverse Wishart distributions, with trivial computational implications. Then, on each sampled parameter set, $\y_t$ is also trivially sampled from the conditional 
compositional model of \eqn{compMVDLMyt}. 
Forecasting to times $t+1,t+2,\ldots$ is similarly performed by easy, direct simulation: predictive Monte Carlo samples are  recursed into the future as in traditional MVDLMs, though now in this more general  setting.

\subsubsection{Conjugate Time $t$ Posterior\label{sec:conjpostt}}

Moving to time $t$, the prior updates to the posterior given new information arising.  

\para{Full Observation on $\y_t$:}  If the full vector $\y_t$ is  observed, the NIW-CNIW 
time $t$ posterior (Theorem~1 of Appendix~\ref{app:CNIW} adapted to the time $t$-specific setting) is 
$$
      p(\bTheta_{ct},\bSigma_{ct}|\cD_t) 
         p(\bTheta_{et},\bGamma_{et}, \bPsi_{et} | \bTheta_{ct},\cD_t)
$$
with components
\begin{equation}\label{eq:CNIWtimetposterior}
\begin{split}
\bTheta_{ct},\bSigma_{ct}|\cD_t &\sim \tNIW(\M_{ct},\C_t,n_t,\D_t),\\
\bTheta_{et},\bGamma_{et}, \bPsi_{et}| \bTheta_{ct}, \cD_t  &\sim  \tCNIW(\Z_t, \C_{et}, \cniwdf_{et}, \H_t | \bTheta_{ct}).
\end{split}
\end{equation}
Updated parameters follow from the general NIW-CNIW theory in
Appendix~\ref{app:p2pNIW-CNIW}, simply now indexed by time $t$.  Prior-to-posterior updates are as follows: 
\begin{itemize} \itemsep-3pt
\item $\M_{ct},\C_{ct},n_t,\D_t$ are defined by the standard updating equations  in MVDLMs-- as in Section~\ref{sec:Bayes_mvdlm} but here applied to the $c-$component of the model and prior.
\item Define the $q-$vector $\z_t$ via  $\z_t'= \y_t'-\F_t'\Z_t^*$ and the $p-$vector 
$\A_{et} = \C_{et}^*\F_t/v_{et}$ with $v_{et} = 1+\F_t'\C_{et}^*\F_t.$   
Then the posterior CNIW parameters are given by 
$\Z_t = \Z_t^*+\A_{et}\z_t'$,     $\C_{et} = \C_{et}^* - \A_{et}\A_{et}'v_{et}$,  
    $s_{et}=s_{et}^*+1$ and $\H_t = \H_t^* + \z_t\z_t'/v_{et}$.
\end{itemize} 
Quantified posterior inferences on model parameters at time $t$ are trivially evaluated by direct simulation of this posterior.  The compositional form is simulated by Monte Carlo draws from $p(\bTheta_{ct},\bSigma_{ct}|\cD_t)$ followed by draws from 
$p(\bTheta_{et},\bGamma_{et}, \bPsi_{et}| \bTheta_{ct}, \cD_t )$ at each sampled value of $\bTheta_{ct}$.

\para{Observation on $\y_{ct}$ alone:}  Of key relevance to the causal counterfactual setting, suppose $\y_{ct}$ is observed but $\y_{et}$ is unobserved, regarded as missing at random.  The update proceeds as usual for the $c-$marginal parameters $(\bTheta_{ct},\bSigma_{ct}).$   However, the time $t$ posterior for the $e|c-$parameters $(\bTheta_{et},\bGamma_{et}, \bPsi_{et})$ is simply the time $t$ prior.

\subsubsection{Decouple/Recouple and MVDLM Special Case} 

A key feature is that the $c-$marginal component of the model is \lq\lq conditionally decoupled'' from that for the $e|c-$conditional, with 
$\bTheta_{ct}$ inducing the decoupling. The overall analysis \lq\lq recouples'' across the $c-$ and $e|c-$components for inference and prediction,  representing an instantiation of the decouple/recouple modelling concept~\citep{West2020Akaike}. 
It is also to be stressed that the standard MVDLM analysis arises as a special case.  
Specifically, if $\beta_e=\beta,\delta_e=\delta$ and analysis begins with a time $t=0$ NIW prior,  the traditional MVDLM analysis is recovered over time periods $t>0$ during which full observations on $\y_t'=(\y_{ct}',\y_{et}')$ are made.  

\subsection{Causal Inference and Prediction} 

\subsubsection{Counterfactual Dynamic Models and Causal Prediction\label{sec:predictiveandfilteredcausal} }
In the counterfactual setting of Section~\ref{sec:causalsetting},  $\y_{ct}$ is designated as the vector control series and-- in our notation so far-- $\y_{et}$ the experimental.  Up to time $t=T$,  they are jointly observed and thereafter the latter is \lq\lq missing''-- the hypothetical latent counterfactual series.  
The explicit notation reflects this:  $\y_{e_0t}$ is the hypothetical counterfactual series and $\y_{e_1t}$ is the actual outcome series. That is, 
$\y_t'=(\y_{ct}',\y_{e_1t}')$ for $t<T$ and $\y_t'=(\y_{ct}',\y_{e_0t}')$. In the  general theory of Section~\ref{sec:compositionalMVDLMs}, this
simply relabels the $e$ series according to whether and when it is observed or latent.  Then, $\y_{e_1t}$ is, of course, actually observed for $t\ge t$;   this standard counterfactual framework ignores those outcomes in the model analysis post-intervention. 
Investigating (in)consistency of the actually realized outcomes $\y_{e_1t}$ with inferences and predictions from the counterfactual model is the primary causal focus.  A compositional model of Section~\ref{sec:compositionalMVDLMs} applies over all time. The context is consistent with treating the $c,e$ series exchangeably prior to the intervention time $T.$ The standard MVDLM analysis applies for $t<T$ with discount factors 
$\beta_e=\beta,~\delta_e=\delta$ and a specified NIW prior at $t=0.$   For $t\ge T$ the evolution/predict/update steps are based on the more general 
NIW-CNIW theory of compositional models. In order that resulting counterfactual inferences be related solely to the missingness of $\y_{e_0t}$, the same model structure  and discount factors are maintained post-intervention.

Post-intervention, analysis generates sequentially revised forecast distributions $p(\y_{e_0t}|\cD_{t-1})$ yielding $1-$step ahead inferences on the {\em  predictive causal effect} $\y_{e_1t}-\y_{e_0t}.$   This recognizes full uncertainty in the predictive setting.  Then, 
on observation of $\y_{ct}$, this can be trivially updated to time $t$ filtered causal inference: generate a direct Monte Carlo sample for the 
posterior predictive distribution of $\y_{e_1t}-\y_{e_0t}$ by draws from 
model parameters from the time $t$ posterior of Section~\ref{sec:conjpostt} followed by draws of $\y_{e_0t}$ from the
$e-$model component of~\eqn{compMVDLMyt}.
 
In some applications, look-ahead causal predictions are contextually important~\citep[e.g.][]{TierneyEtAl2024}. Here the ability to forecast multiple steps ahead-- with full predictive distributions that properly characterize uncertainties about the future development of $\y_{e_0t+k}$ over $k{=}1,2,\ldots$-- is of practical importance. Compositional MVDLMs enable this by simply recursing simulation-based forecasts into the future. 
 
\subsubsection{Outcome Adaptive Dynamic Models and Causal Prediction} 

In the post-intervention period the actual outcomes $\y_{e_1t}$ are observed. Much of the causal literature generally only uses these outcome, as discussed above, to contrast with the counterfactual predictions. One innovation in the current paper is go further and actively model the post-intervention development of $\y_{e_1t}$-- separately from, and independently of, the counterfactual analysis. The rationale is that understanding  actual post-intervention evolution can and will impact inferences and predictions under the non-counterfactual hypothesis, i.e., admitting and allowing for causal intervention effects on the  experimental series.   The goals are to: (i) define an {\em outcome adaptive model (OAM)} to run in parallel to, and explicitly independent of, the  counterfactual model; (ii) to explore inferences under this OAM relative to the counterfactual model to more properly understand potential causal effects; and (iii) to continually monitor and update the OAM analysis-- and update predictions based on it-- so as to be able to adapt in case a future decision is made  that there have indeed been \lq\lq changes due to intervention''. Following such a decision, inferences and predictions  relative the counterfactual \lq\lq null hypothesis'' can then be accessed.    In applications, such a decision will have implications for moving forward with  future analysis  based on the OAM. 

As an example, suppose a marketing intervention has a large initial effect on sales, but that it then begins to gradually decay over time.  In the short term, the counterfactual evaluations of incoming data highlight the impact of the intervention, but will then be slow to signal the decay of the effect over later times. An appropriate OAM, in contrast, will more rapidly adapt to the decaying effect, being responsive to-- and updated by-- the incoming $\y_{e_1t}$ data. This 
will provide sequentially updated and practically realistic short-term predictions of the continuing decay. 

The OAM defines a parallel model for $\y_{e_1t}$ in the conditional MVDLM framework, in tandem with the counterfactual model for $\y_{e_0t}$ and conditionally independent of $\y_{e0t}.$  The OAM model states and volatilities evolve in time conditionally independently of those in the counterfactual model.    Methodologically, this adds no new questions. The above theory of the sequential filtering and forecasting  in the NIW-CNIW  framework is simply cloned and used 
for series $e_1|c$ in parallel to that for series $e_0|c.$  The independence structure means that observations on the treated experimental outcomes $e_1$ do not at all affect
 inferences and predictions in the counterfactual $e_0|c$ model.  
 
Appropriately treating the $c,e_0,e_1$ series as exchangeable prior to the intervention time $T,$  the MVDLM analysis applies with discount factors 
$\beta_{e_1}=\beta_{e_0}=\beta$ and $\delta_{e_1}=\delta_{e_0}=\delta$.   This is precisely as developed for the $c$ and $e_0|c$ series in the counterfactual analysis, 
now adding the parallel OAM for $e_1|c$.  However, a key novelty is that, at the intervention time $T$ alone, the OAM discount factors $\beta_{e_1},\delta_{e_1}$ are reduced
to smaller values; this model intervention applies only at the specific time $T$ with discount values reverting to the common values $\beta,\delta$ thereafter.  This is a standard statistical response to the view that the intervention {\em may have} a causal impact on the $e_1$ series, and that lower discount factors at that instant will allow for more adaptation in learning under the OAM in the short term~\citep[e.g.][chap.~11]{West1986,West1989,west:harri:97}.
%

\section{A Simulated Example}

\begin{figure}[htbp]
\centering
 \includegraphics[height=6cm]{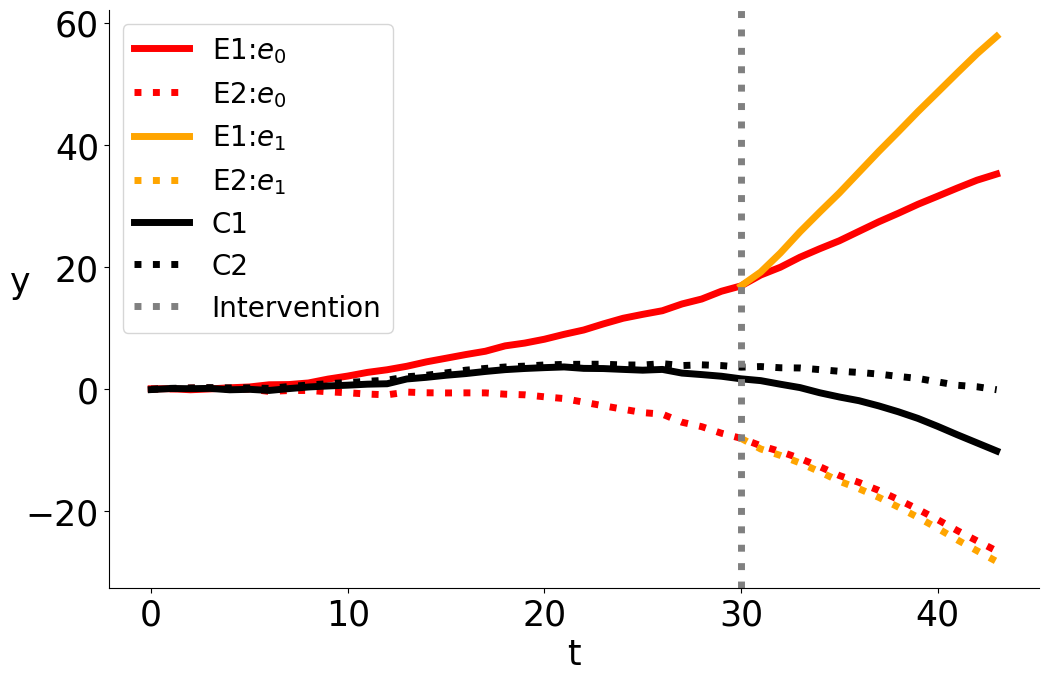}
\qquad\qquad
  \raisebox{1.25cm}{\includegraphics[height=4cm]{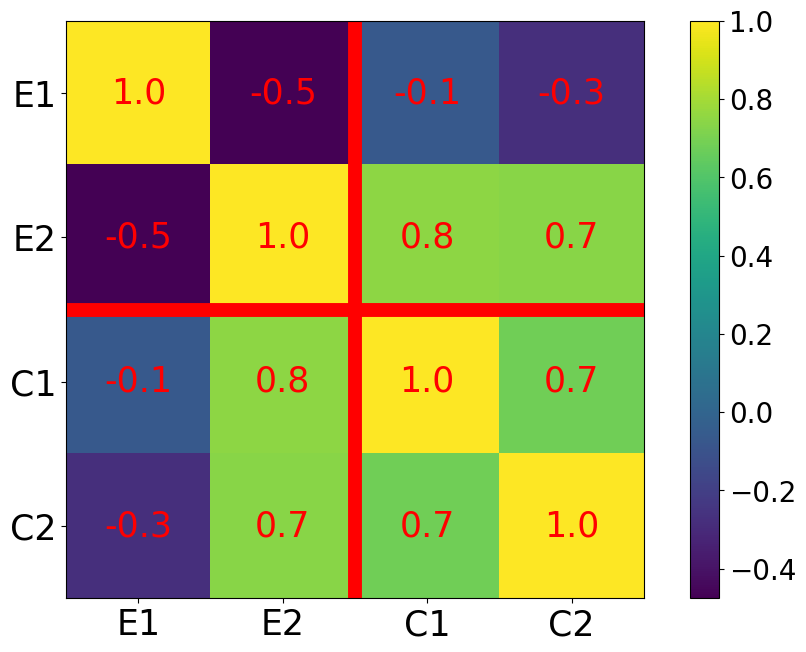}}
\caption{Synthetic  data and underlying cross-series correlation matrix $\R$ in simulated example.}
\label{fig:sim_illustrate}
\end{figure}

To fix ideas and illustrate analysis, a simple, low-noise example uses synthetic data so that we know the true underlying post-intervention counterfactuals $\y_{e_0t}$ as well as the post-intervention actuals $\y_{e_1t}.$  The example uses 
$q = 4$ series with 2 controls (labelled $C1,C2$) and 2 experimental (labelled $E1, E2$).
 The series were generated from a damped linear growth MVDLM with constant $\F_t=\F=(1,0)'$, constant 
$\G_t=\G$ having  first row  $(1,r)$ and second row $(0,r)$ using damping factor $r=0.95$, and a constant cross-series variance matrix $\bSigma_t=\bSigma$ drawn from 
$\tIW(4,\R)$ with a specified $\R.$  
Intervention at  $T=30$ is mimicked by applying a random shock to the state matrix $\bTheta_{eT}$ for the experimental series, with a stronger impact on 
$E1$ than on $E2$.  Fig.~\ref{fig:sim_illustrate} shows the series  
and the correlations in the realized $\bSigma.$ Note that $E1$ has much weaker pre-intervention correlations with controls than does $E2$.
 
\begin{figure}[htbp]
\centering
\subfloat[]{%
  \includegraphics[width=15cm]{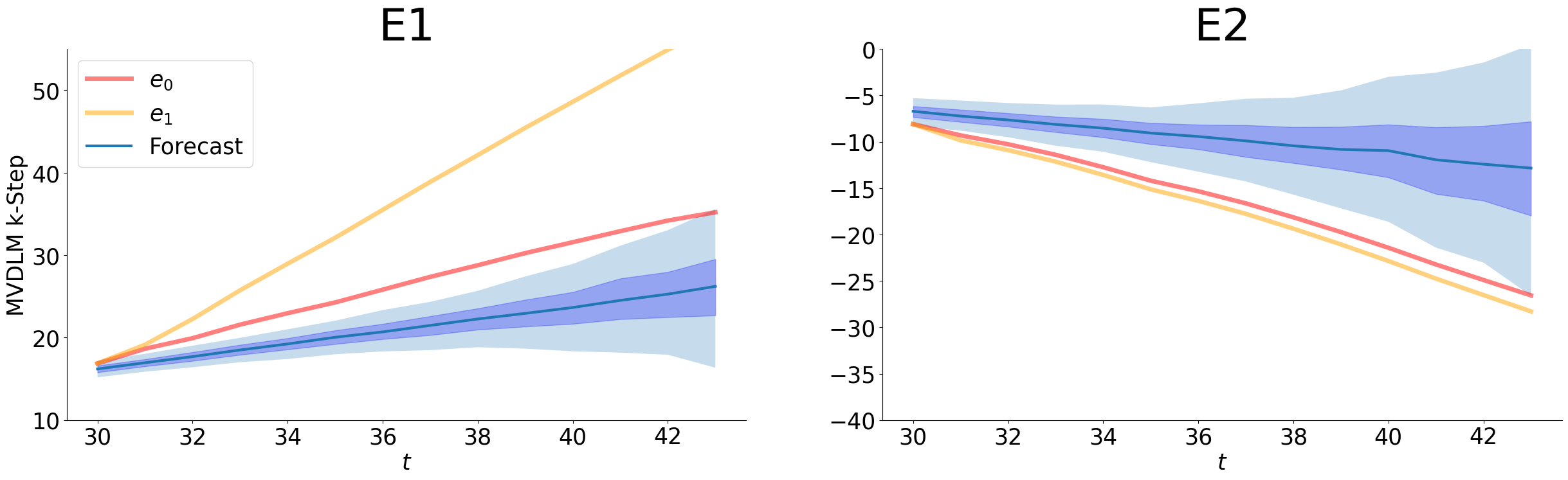}%
  \label{fig:sim_result:MVDLM}%
} \\
\subfloat[]{%
  \includegraphics[width=15cm]{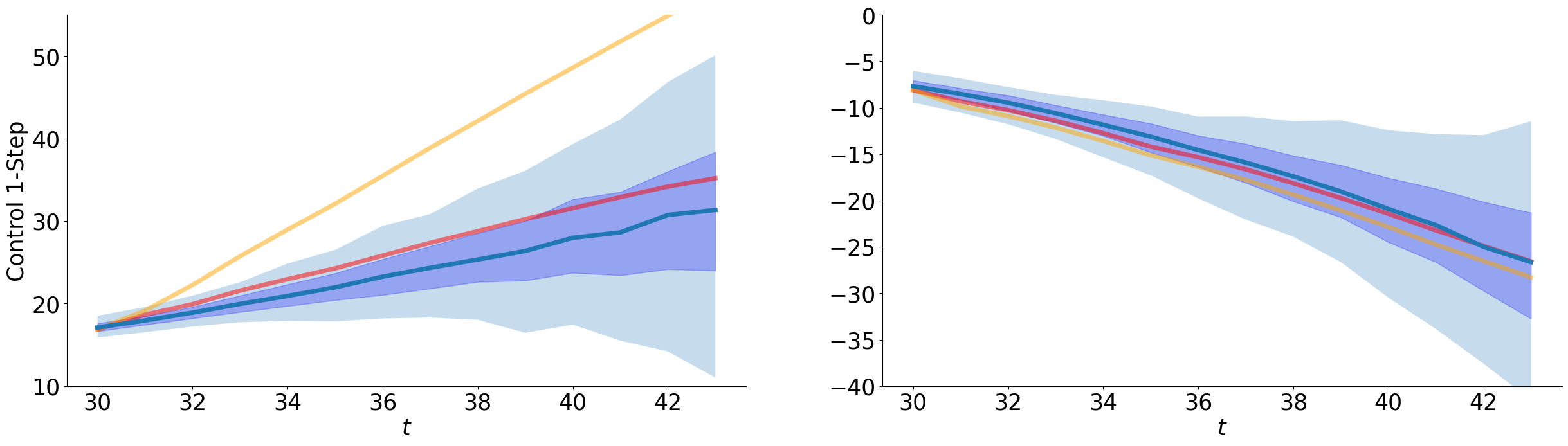}%
  \label{fig:sim_result:SCM}%
} \\
\subfloat[]{%
  \includegraphics[width=15cm]{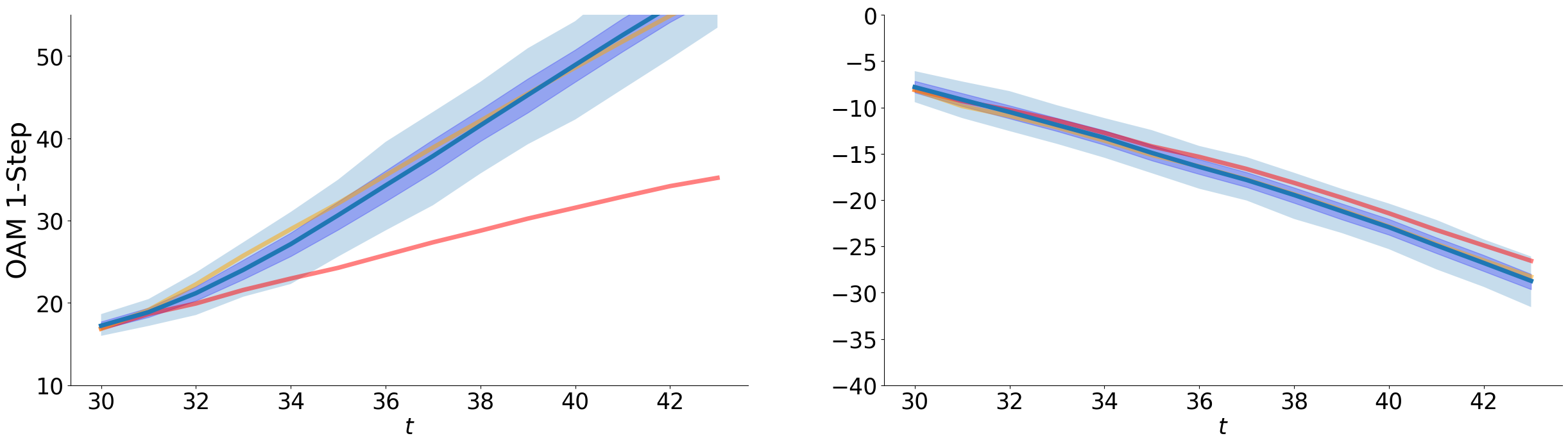}%
  \label{fig:sim_result:OAM}%
}
\caption{Simulation example model predictions. (\ref{fig:sim_result:MVDLM}): Multi-step forecasts made at $t=29$ over the following intervention period, with the true $\y_{e_0t}$ and actual outcome $\y_{e_1t}$ series.   
(\ref{fig:sim_result:SCM}): Sequentially revised counterfactual $1$-step forecasts; (\ref{fig:sim_result:OAM}): $1$-step forecasts from the  OAM. Forecasts are shown in terms of medians, $50\%$ credible intervals (darker) and $90\%$ credible intervals (lighter). 
}
\label{fig:sim_result}
\end{figure}

\begin{figure}
    \centering
    \includegraphics[width = 0.95\textwidth]{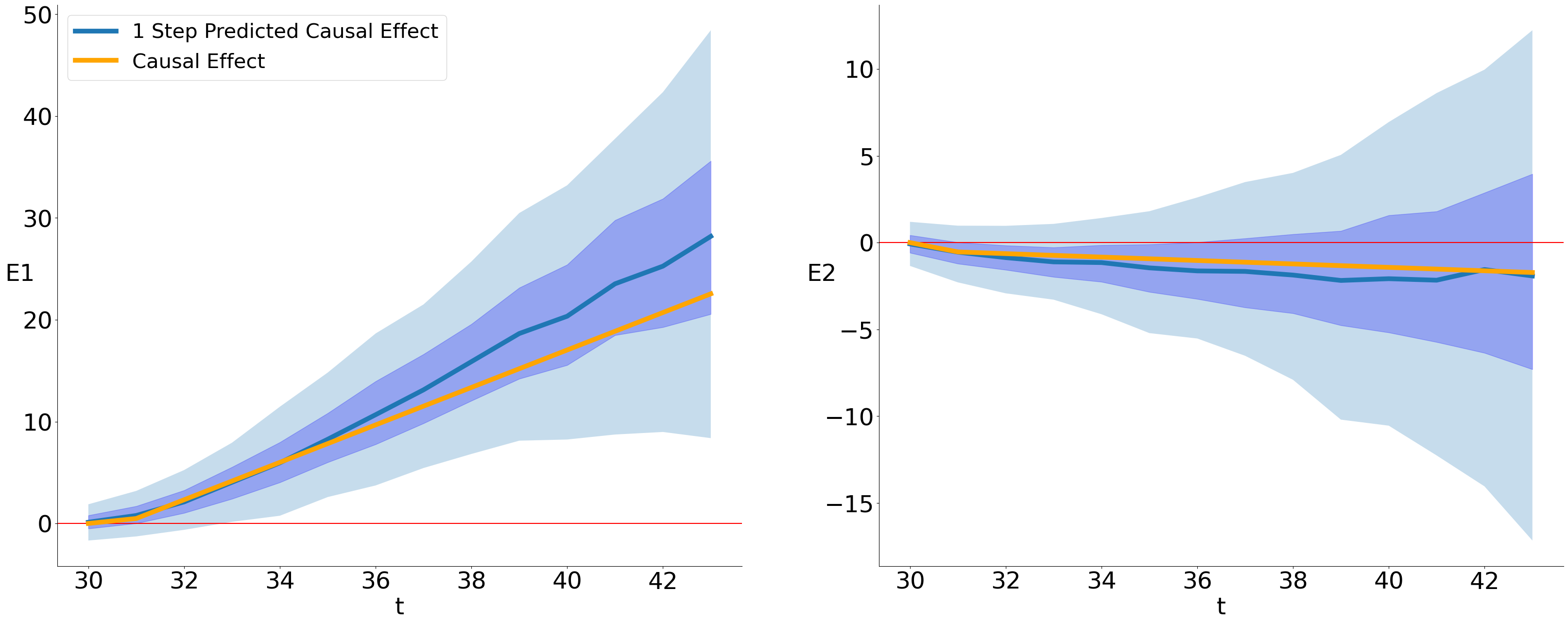}
    \caption{Simulation example causal prediction showing sequentially revised $1-$step predictions for E1 and E2.   Forecasts are shown in terms of medians, $50\%$ credible intervals (darker) and $90\%$ credible intervals (lighter). } 
    \label{fig:sim_effect}
\end{figure}

Fig.~\ref{fig:sim_result} displays predictions from the compositional MVDLM analysis.  
Fig.~\ref{fig:sim_result:MVDLM} shows $k-$step ahead predictions made at time $t = 29$ over  $k=1,2,\ldots$ into the post-intervention period.  Fig.~\ref{fig:sim_result:SCM} shows sequentially updated $1-$step predictions under the counterfactual, and Fig.~\ref{fig:sim_result:OAM} those in the OAM. We see the unconditioned MVDLM forecast simply continues to predict the  trend estimated from the pre-intervention period. The conditional synthetic control model regularizes the MVDLM forecasts towards the trend of the controls. The inferred correlation at the end of the pre-intervention period influences the nature of this regularization. The synthetic control forecasts for the $E1$ series lie between the untreated outcome and the unconditioned MVDLM forecast due to the weak correlation with the control series. In contrast, the  control forecasts for $E2$ match the untreated outcomes  closely due to the high correlation with the control series. This result is notable, as the rate of change for $E2$ increases notably in the post intervention time period, as evinced by the poor predictions made by the $k-$step ahead MVDLM forecast. Further, the OAM inflation of uncertainty at $T=30$-- based on lower discount factors for the $e_1$ model components at only that time-- induces substantial and appropriate adaptation to the 
 dynamics of the treated series over the next few time points. Finally,  Fig.~\ref{fig:sim_effect} shows  the Monte-Carlo based $1-$step ahead predictions of the treatment effect $\y_{e_1t}-\y_{e_0t}$, again sequentially over time; these causal predictions are compared with the actual underlying counterfactual values. The figure shows strong concordance between the causal predictions 
 and the known \lq\lq truth'' in this synthetic example.

\FloatBarrier

\section{Application: Retail Promotions and Revenue Outcomes}

\subsection{Setting and Data} 

Among our areas of motivating applications  are commercial marketing studies.  The example here comes from 
a national retail chain where the intervention is identified as the start of a period of targeted promotion campaigns aimed at historically frequent customers (households).\footnote{The data come from retail data science company \href{https://www.dunnhumby.com/}{dunnhumby}, via the 
\href{https://www.kaggle.com/datasets/frtgnn/dunnhumby-the-complete-journey}{kaggle} public data resource.}
The  data set provides household-level, weekly revenue (spending); this is used to define
a $q=4$-dimensional time series of {\em average weekly revenue} for 4 defined subsets of households, detailed below. 
There are 32 weeks of pre-intervention data, but the first 5 weeks are excluded from analysis due to staggered entry of households into the data set; this period is followed by data for 12 post-intervention weeks to evaluate causal predictions.

\begin{figure}[htbp]
\centering
\subfloat[]{%
  \includegraphics[width=10cm]{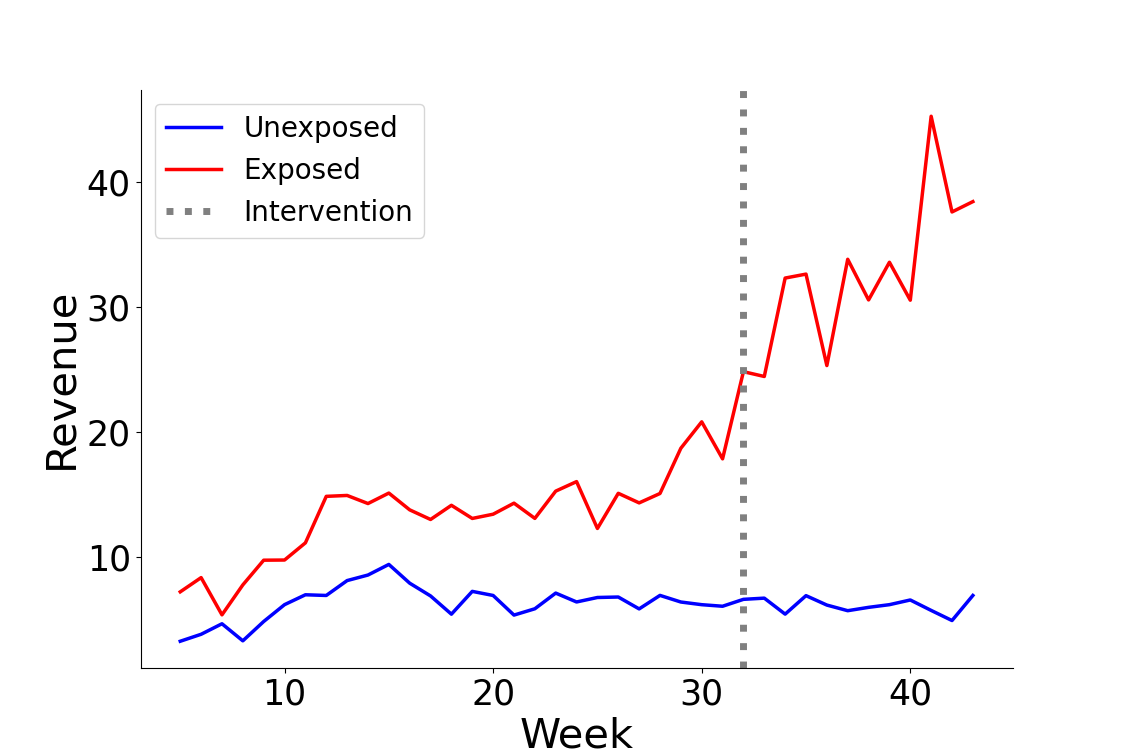}%
  \quad 
  \raisebox{1.5cm}{\includegraphics[width = 6cm]{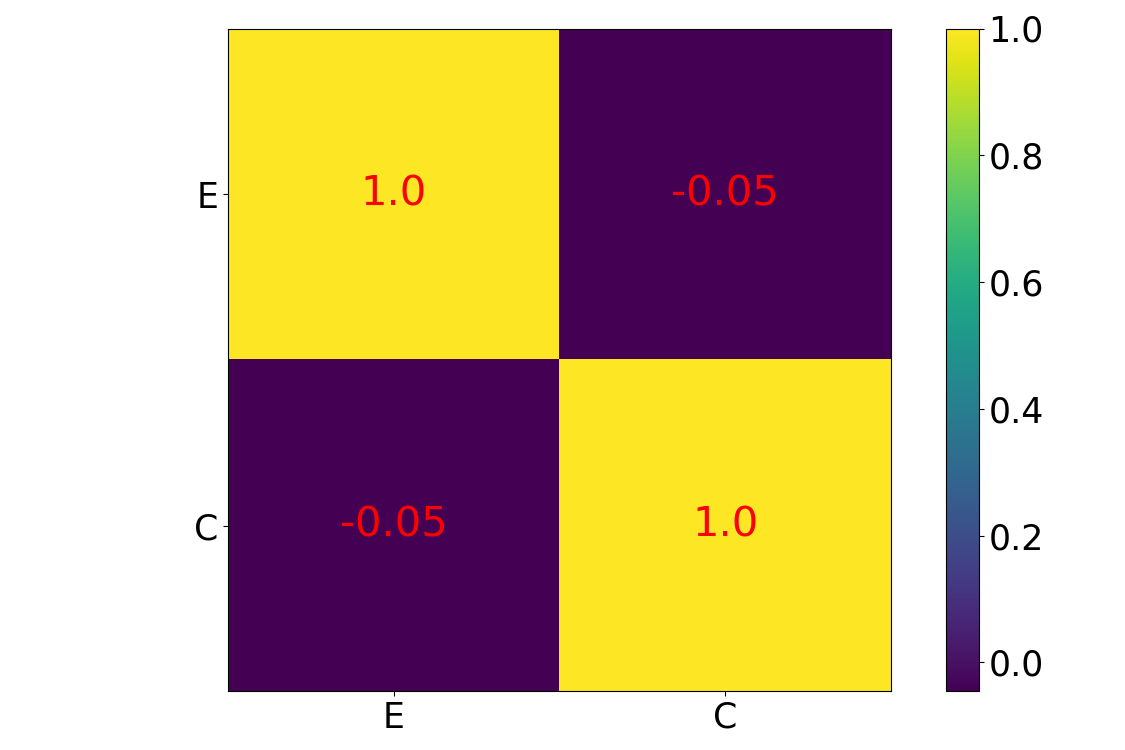}}%
  \label{fig:dh_data:summary_2}%
  }
\\
\subfloat[]{%
  \includegraphics[width=10cm]{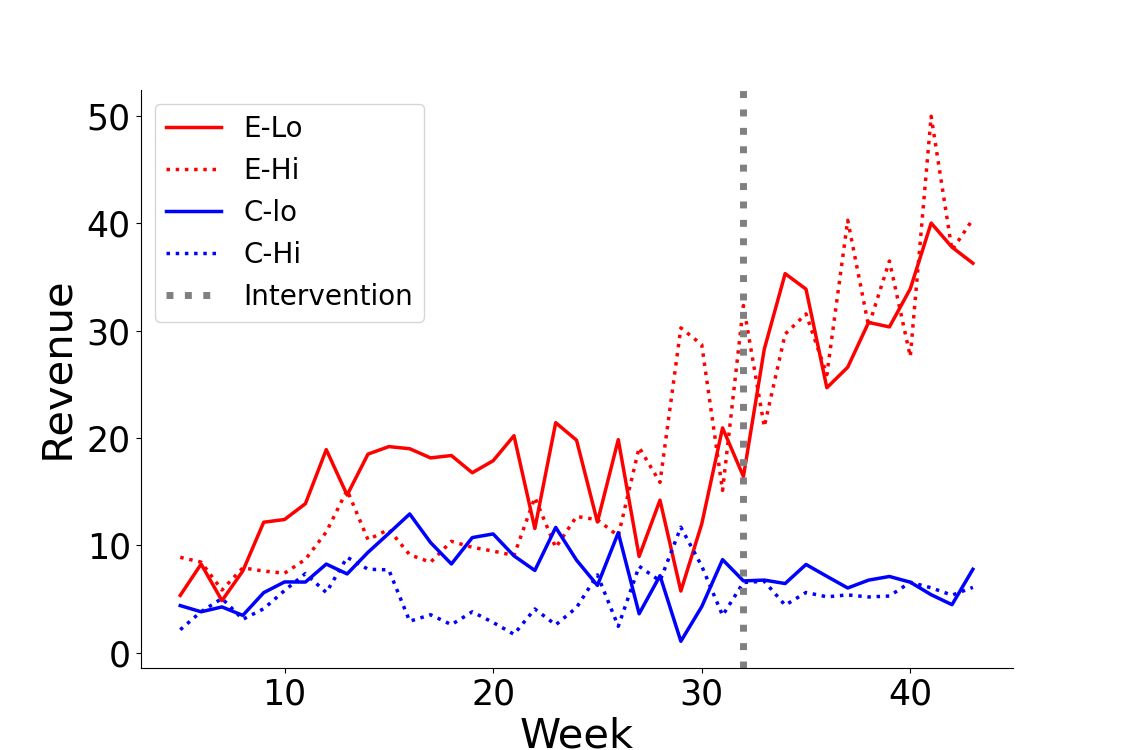}
  \quad 
  \raisebox{1.5cm}{\includegraphics[width = 6cm]{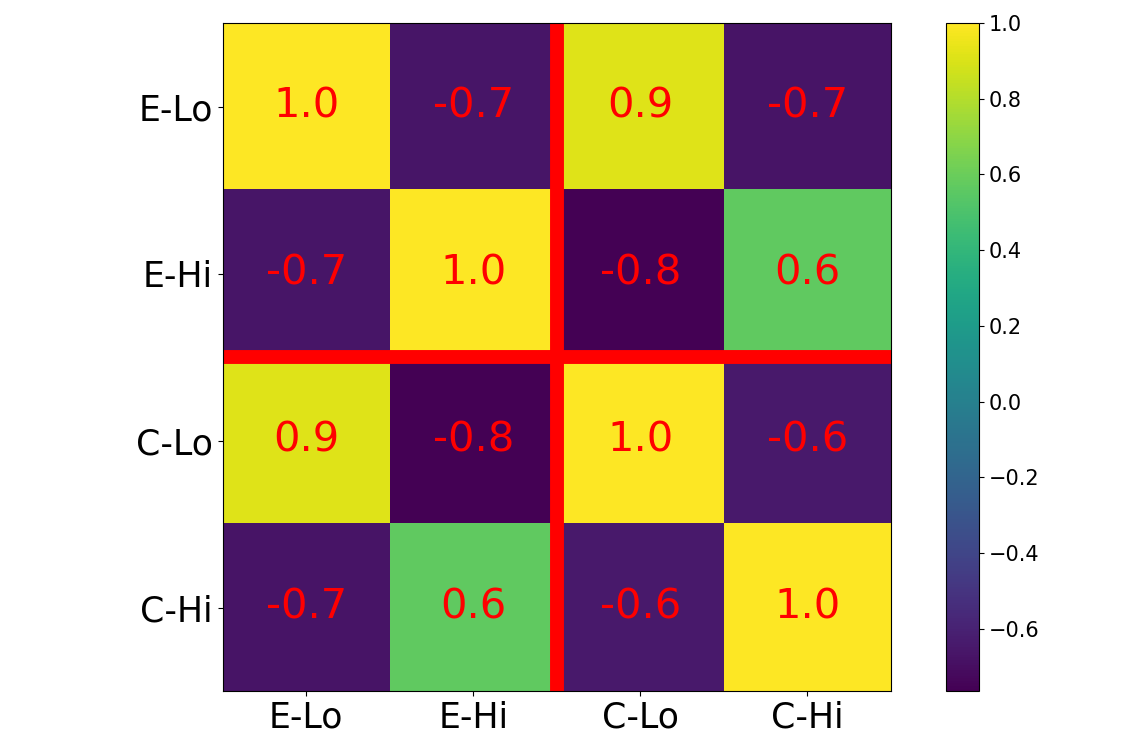}}%
  \label{fig:dh_data:summary_4}%
  }
\caption{Time series of average weekly revenue in the retail study. 
(\ref{fig:dh_data:summary_2}): Revenue for “exposed” and “unexposed” households from weeks $t = 5$ to $t = 44$, and
pre-intervention sample correlations. (\ref{fig:dh_data:summary_4}): Revenue for households in the 4 groups defined by  
exposed/unexposed to promotion crossed with Hi/Lo pre-intervention factor, \lq\lq exposed'' now labelled \lq\lq C'' for control, and unexposed \lq\lq E'' for experimental;  pre-intervention sample correlations evidence strong dependencies.}
\label{fig:dh_summary}
\end{figure}

The focus is on a subset of 956 households that were relatively low spending (average weekly revenue less than \$25)  
during the pre-intervention period. Among these, we identify two groups:  731 with no known exposure to the 
promotions, and 225  known to have received a high level of promotions. The average revenue series in these groups are shown in Fig.~\ref{fig:dh_data:summary_2}.

The pre-intervention sample correlation between the exposed/unexposed series control series is negligible $(\sim -0.05)$. This is not surprising as the two groups are likely composed of heterogeneous households that follow differing dynamics.  Then, a key  and important feature is that the exposed group has higher average revenue than the unexposed.  The promotion campaign household targeting was not random, aiming to lift revenue  among the more highly spending of the designated households. This non-random assignment is not a concern for our analysis so long as we define appropriate synthetic control series that are related to the designated experimental series over the pre-intervention period. 

We expect that the promotional campaign may have varying effects within different subgroups of the exposed  households, suggesting stratification of households to help define synthetic controls.  This must be based only on pre-intervention data and information.   On the pre-intervention data alone, a singular factor decomposition of the full household $\times$ week data  generates empirical factors (a.k.a. principal components) to explore pre-intervention heterogeneity.  The first factor simply represents average revenue over the weeks with little dependence across groups pre-intervention. The second factor, however, shows substantial pre-intervention dependence across the groups that were designated as exposed/unexposed.   Each of the 956 households has an empirical loading on this factor reflecting variation across households in this factor dimension; we use these loadings to define the  \lq\lq Hi'' and \lq\lq Lo'' household groups according to whether the loading is above or below the median loading across all households.   Again, this is based purely on pre-intervention data.  This defines $q=4$ groups by crossing the exposed/unexposed with Hi/Lo classifications; average weekly revenue of these groups in Fig.~\ref{fig:dh_data:summary_4} shows evident differences in pre-intervention dynamics related to the Hi/Lo categorization.  Importantly, pre-intervention sample correlations are all high (in absolute value). In particular, 
strong pre-intervention dependence of  exposure and factor classifications suggest
 2 control series $\y_{ct}$ and   2 experimental series $\y_{et}:$  control series C-Hi, C-Lo  (unexposed/factor Hi, unexposed/factor Lo), 
 and experimental series  E-Hi, E-Lo  (exposed/factor Hi, exposed/factor Lo).

\subsection{Compositional MVDLM and Causal Prediction Analysis} 
The specified MVDLM applies to logged average revenue in each of the $q=4$ groups.  This has a damped linear growth form with constant $\F_t=\F=(1,0)'$, constant 
$\G_t=\G$ having  first row  $(1,r)$ and second row $(0,r)$ based on damping factor $r=0.95$.  Discount factors  $\delta =0.8, \beta = 0.95$ underlie the evolutions of the matrices and volatility matrices, respectively; these allow for more stochastic variation over time in each series-specific trend than in residual volatilities and cross-series correlations. At the intervention week $T=32$, the discount factors in the OAM are momentarily dropped to 
$\beta_{e_1}=0.85,\delta_{e_1}=0.7$ anticipating potential post-intervention changes.  
Initialization at $t=6$ (on the original weekly time scale) uses a relatively vague prior with
 $\C_6^*=5 \I$, $n_6^*= 10$ and $\D_6^*= \I$. For $\M_6^{*}$ we set the first row to be the average value of the series over $t = 1,\ldots,5$ and the second row to be zero. Through the analysis period, forecast outcomes are translated to the revenue sale for summaries. 
 
\begin{figure}[htbp]
\centering
\subfloat[]{%
  \includegraphics[width=9.5cm]{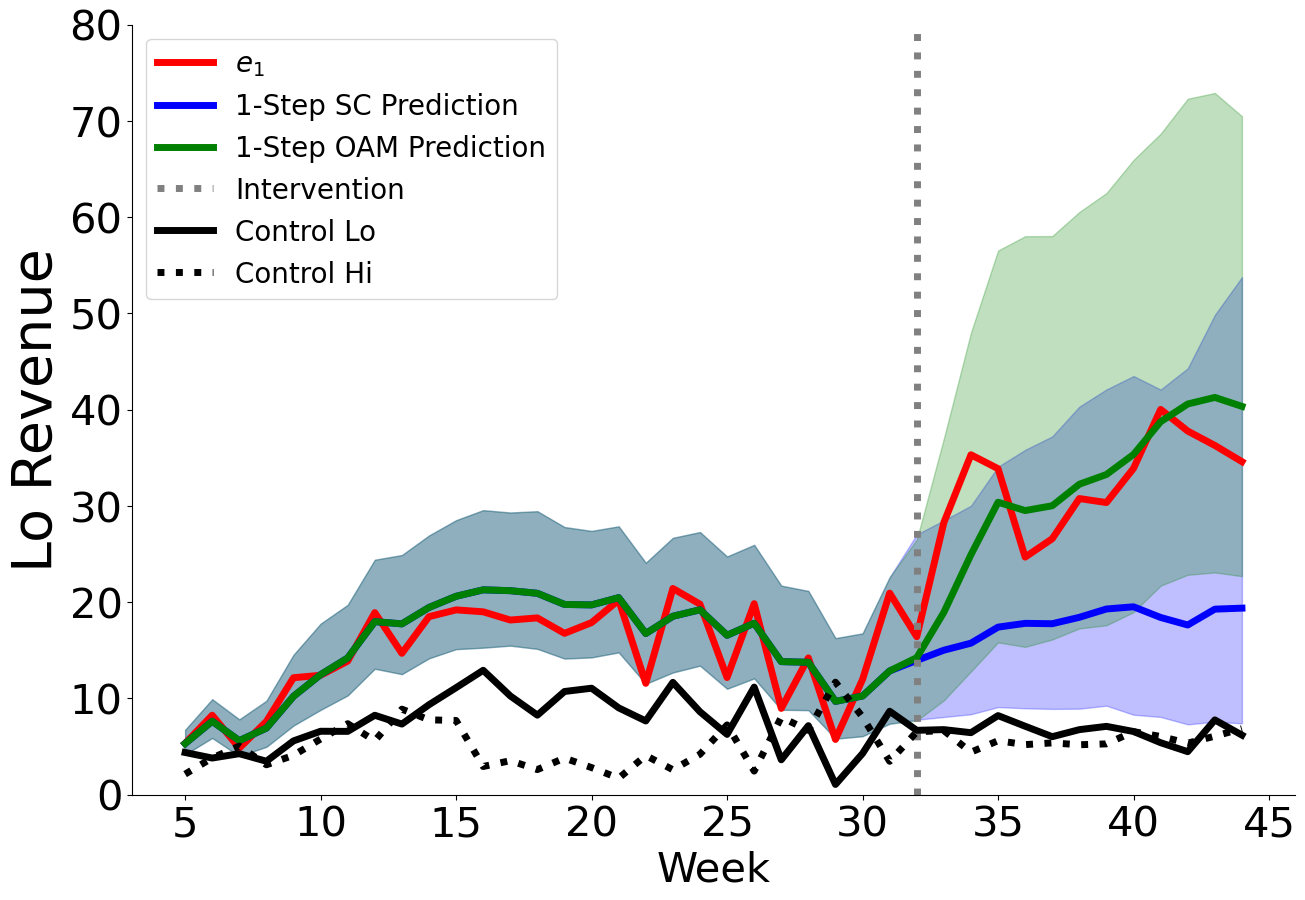}%
  \label{fig:dh_data:pclo}%
} \ \ \ 
\subfloat[]{%
  \includegraphics[width=9.5cm]{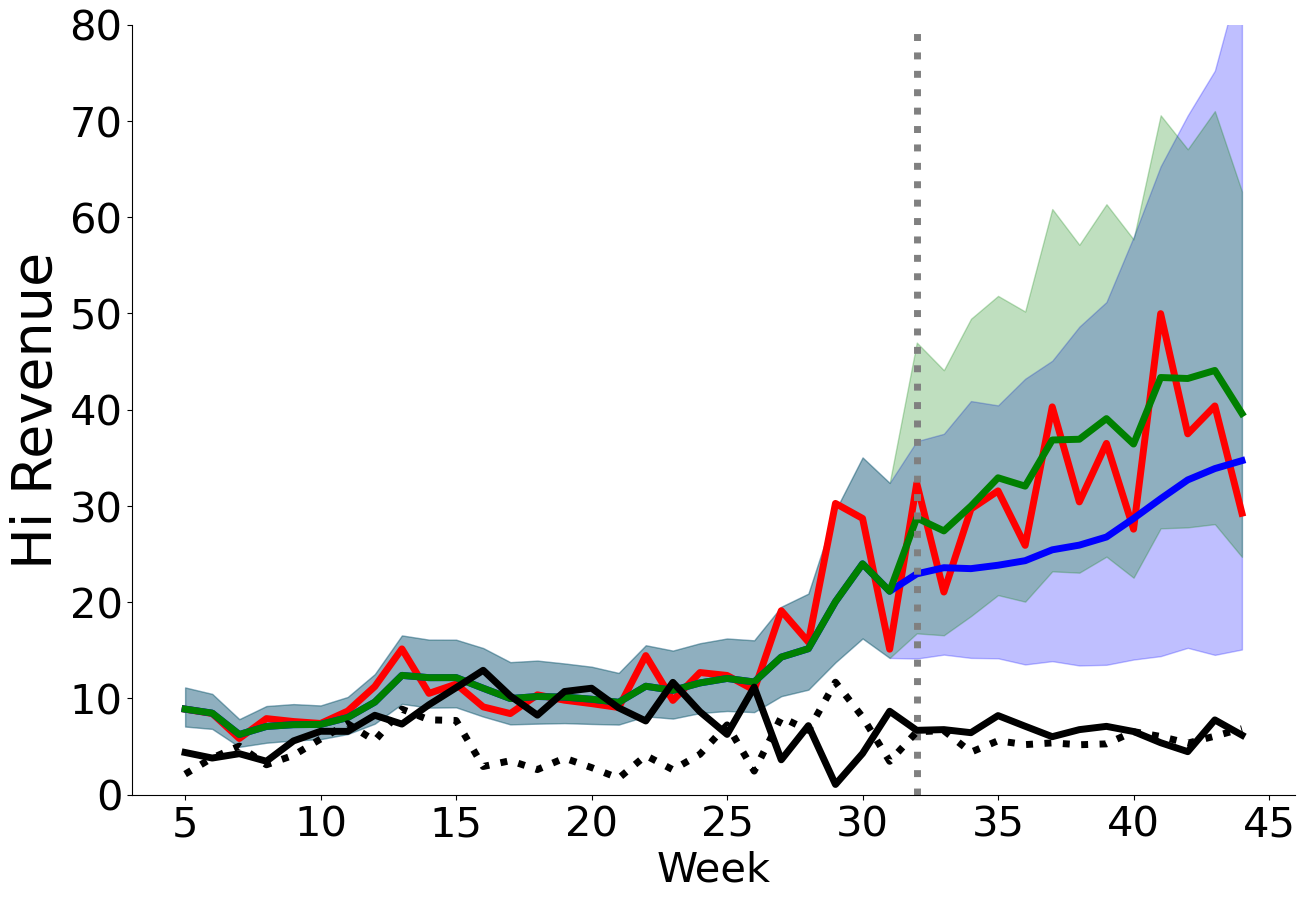}%
  \label{fig:dh_data:pchi}%
}
\\
\subfloat[]{%
  \includegraphics[width=5cm]{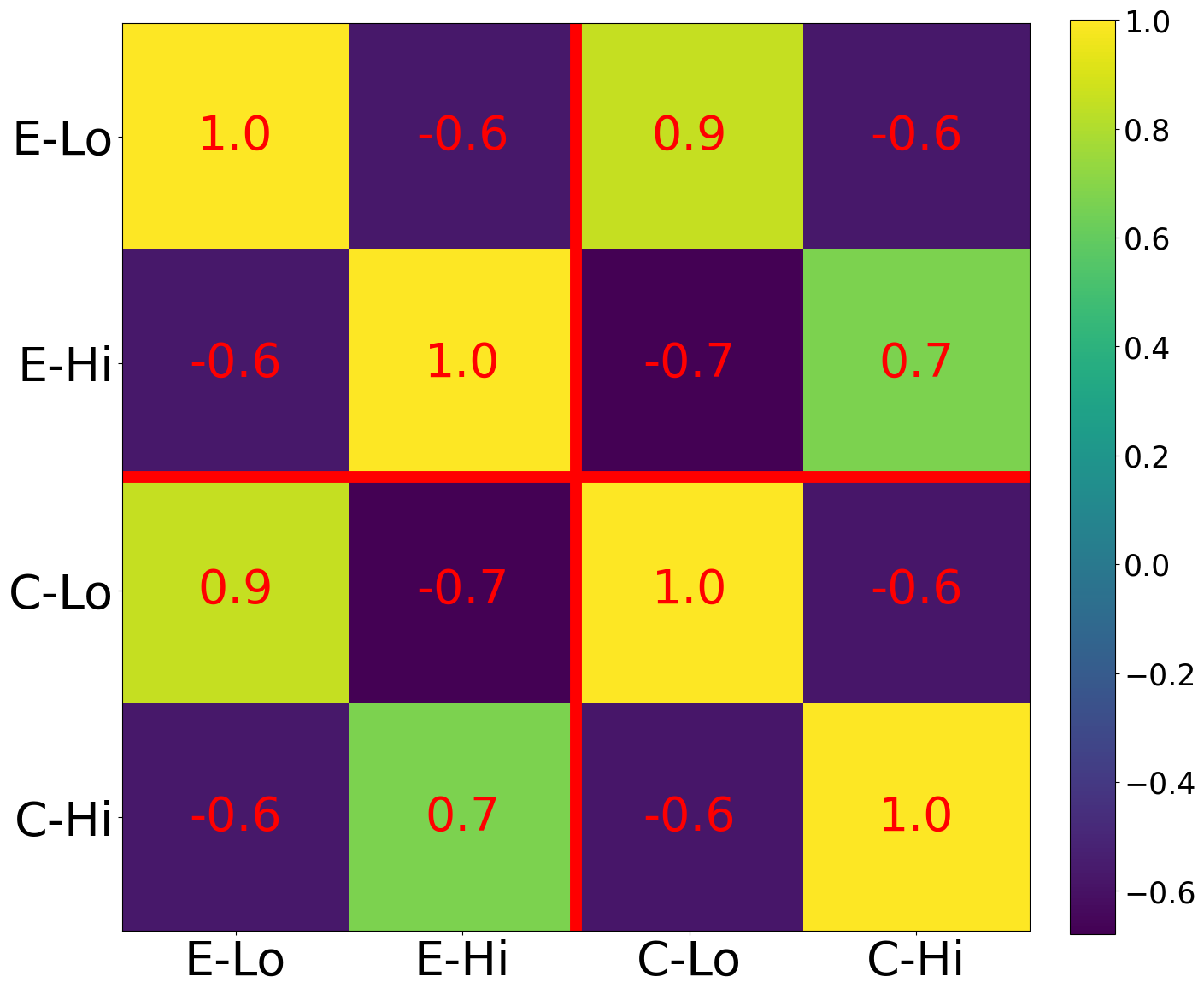}%
  \label{fig:dh_data:est_cor}%
} 
\caption{Weekly average revenue, $1-$week ahead forecasts in the retail study. Forecasts are from the counterfactual model and the OAM  in each of the 
 Lo (\ref{fig:dh_data:pclo})  and  Hi  (\ref{fig:dh_data:pchi})  groups. The format follows earlier figures with medians and 50\%, 90\% intervals.  The prior predictive correlation matrix at intervention time $T= 32$ (\ref{fig:dh_data:est_cor}) shows strong pre-intervention cross-series dependencies. }
\label{fig:dh_pclopchiestcor}
\end{figure}

Fig.~\ref{fig:dh_data:pclo} and \ref{fig:dh_data:pchi} shows resulting $1-$week ahead predictions of average revenue for the Lo and Hi series, respectively.
Predictions show median and 50\%,90\% forecast intervals from the counterfactual and the OAM components.
Conditioning on the control series somewhat tempers the forecast growth of the counterfactual in the Lo group,   while it serves to somewhat 
increase the forecast growth in the Hi group.  Predictions in the OAM closely track the actual treated outcomes  with increased 
 uncertainty immediately post-intervention; this is partly due to the use at the intervention week of lower discount factors, and the OAM then adapts more 
 appropriately to higher levels of variation in both trend and  volatility in the experimental groups post-intervention. 
Fig.~\ref{fig:dh_data:est_cor} displays estimated correlations from the prior for $\bSigma_T$, i.e., the correlations evaluated from $\D_T^*$ immediately prior the intervention. These compare closely with the purely empirical sample correlations noted earlier in Fig.~\ref{fig:dh_data:summary_4}.

%

\begin{figure}
    \centering
    \includegraphics[width = 0.95\textwidth]{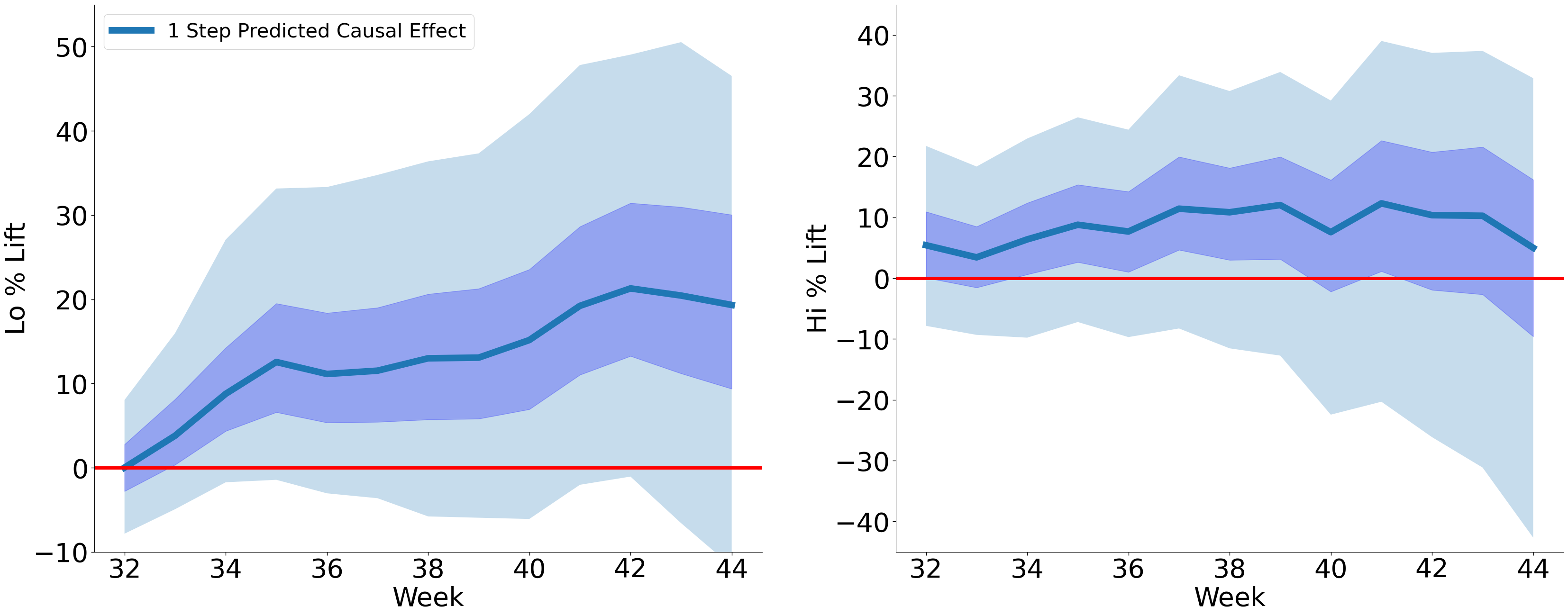}
    \caption{Causal prediction of intervention effects in the retail study. This is shown in terms of $1-$week ahead forecasts of \% lift for the Lo and Hi groups; the format follows earlier figures with medians and 50\%, 90\% intervals. }
    \label{fig:dh_forecastlift}
\end{figure}

\newpage

%
%

The Monte Carlo samples from $1-$week ahead forecasts of the synthetic controls underlie summaries of causal analysis in Fig.~\ref{fig:dh_forecastlift}.
As the model is for revenue on the log scale, the transformed causal effects $100 \{\exp(\y_{e_1t}-\y_{e_0t})-1\}$ define weekly percent \lq\lq lift in revenue'' 
predicted due to the intervention. The figure shows the sequentially updated $1-$step predictions for \% lift for each of the Lo and Hi groups.  The 90\% credible intervals for the Lo group barely includes zero,  while 
the 50\% interval is solidly within the positive lift region throughout the post-intervention time period. Lift in the Hi group is generally forecast at lower values and apparently less statistically significant.  That said, a forecast of lift with a better than 75\% probability of exceeding zero could be emphasized as of potential practical significance, depending on the follow-on decision analysis that such causal predictions feed into. 

The above analysis focused exclusively on causal prediction with the $1-$week ahead inferences that
recognizes full uncertainty in the predictive setting.  As noted discussed in Section~\ref{sec:predictiveandfilteredcausal}, we can go further at each time $t$ to 
evaluate the causal analysis in the filtered sense,  updating the $1-$step \lq\lq prior'' on the causal effects to be based on the time $t$ posterior having observed 
$\y_{ct}$. This can make a difference in terms of location and uncertainty. In the current case study, the differences are negligible, mainly due to the high levels of volatility in post-intervention outcomes.
The posterior lift trajectories (not shown) are practically closely similar to the $1-$step trajectories in 
Fig.~\ref{fig:dh_forecastlift}, so follow-on decisions would not be materially impacted. This may or may not be the case, of course, in other applications where comparison of forecast and fitted inferences on causal effects may have more practical import.  

\FloatBarrier

\newpage
\section{Summary Comments\label{sec:conc}}

Key aspects of the compositional MVDLM framework include characterising cross-series dependencies among controls and experimental series 
pre-intervention, and allowing for such dependencies to change over time. Then, causal effects may change over the post-intervention period,  and it is important to properly communicate corresponding  counterfactual uncertainties.  A fully Bayesian multivariate model that is amenable to direct evaluation of inferences and predictions on primary counterfactual series as well as functions of them is desirable in many applications, as is sequential and time-adaptive analysis for monitoring post-intervention data. We also stress access to  theoretically precise and computationally scalable   methodology.  These goals are addressed in this paper, with methodology that advances the utility of MVDLMs in causal settings. The theoretical development exploits compositional structure and represents an example of a decouple/recouple modelling strategy~\citep{West2020Akaike}.  The methodology defines computationally simple, scalable Bayesian analysis for counterfactual causal inference and prediction using any number of designated series as synthetic controls.   Additional developments include the use of the outcome adaptive model for post-intervention monitoring for potential use in resulting decisions about the practical relevance of inferred causal effects. 

There are various potential future directions for both applying compositional MVDLMs and for new methodological extensions.   Applications in various areas will require specification of series to be used as synthetic controls. In more highly multivariate settings this can build on the example here in using exploratory analysis of pre-intervention data to explore and exploit heterogeneity among series.   Then, extension to causal settings where experimental series are enrolled in a study sequentially-- the settings of staggered  \lq\lq roll-out'', a.k.a. staggered accrual of experimental series into the intervention study-- is one key interest. The structure of compositional MVDLMs is amenable to such extensions, but details and questions of scalability are open for investigation.    The MVDLM framework admits many choices of model structure through the $\F_t,\G_t$ components that future applications may exploit. Some such choices lead to time-varying vector 
autoregressions, workhorses in traditional analysis and forecasting econometrics and allied areas. Such models raise questions about extension to the causal setting. Autoregressions require recent values of some of the series used as predictors of others for the near-term future; the challenge then arises if some of these predictor series become experimental post-intervention in models for the continuing control series. Related  
questions arise in settings where there is interest in alternatives to the exchangeable structure of MVDLMs with respect to both the specifications of $(\F_t,\G_t)$ and to cross-series dependencies. Relaxing these assumptions-- such as towards simultaneous graphical 
dynamic linear models~\citep{GruberWest2016,GruberWest2017}-- will modify the framework, and open up questions of how the the compositional perspective and theory 
for causal prediction will be developed.

\newpage

\begin{appendices}
	
\section{Inverse Wishart Notation and Summaries}

\paragraph{Terms:}  
\begin{tabular}{ll} probability density function: p.d.f. & symmetric positive definite: s.p.d.  
  \\ degrees of freedom: d.o.f.  & $\etr(\A): \ \exp\{\textrm{trace}(\A)\}$ for any matrix $\A$\\
   Wishart: $\tW$ & inverse Wishart: $\tIW$ \\ (matrix) normal-inverse Wishart: $\tNIW$ & conditional normal-inverse Wishart: $\tCNIW$
    \end{tabular} 
   
\subsection{Inverse Wishart and Wishart\label{app:IW}} 
The $q\times q$ s.p.d. variance matrix $\bSigma\sim\tIW(n,\D)$ has p.d.f. 
$$
 p(\bSigma|n,\D)  \propto  |\D|^{(n+q-1)/2}  |\bSigma|^{-(q+n/2)} \etr (-\bSigma^{-1} \D/2),
$$ 
with {\em inverse Wishart d.o.f.} $n>0,$ s.p.d. $q\times q$ scale matrix $\D$ and $\tE(\bSigma) = \D/(n-2)$ if $n>2.$  
The precision matrix $\bSigma^{-1} \sim\tW(d,\D^{-1})$ with {\em Wishart d.o.f.} $d=n+q-1 >q-1$ and harmonic mean 
$\tE(\bSigma^{-1})^{-1}= \D/d.$  A traditional point estimate of $\bSigma$ is $\S=\D/n.$ 

\subsection{Matrix Beta Stochastic Volatility\label{app:MatrixBeta}}
The matrix beta evolution  $\bSigma_t=\tMB_t(\bSigma_{t-1})$ underlies discount factor-based, conjugate learning of volatility matrices 
(\citealp{Quintana1987}; \citealp{uhlig94}; \citealp[][sect.~16.4]{west:harri:97}; \citealp[][sect.~10.4]{PradoFerreiraWest2021}).    At time $t-1,$ the current posterior has
$(\bSigma_{t-1}|\cD_{t-1})  \sim \tIW(n_{t-1},\D_{t-1})$ 
with harmonic mean
$\tE(\bSigma_t^{-1}|\cD_{t-1})^{-1} = \D_{t-1}/(n_{t-1}+q-1).$ 

Markov transition to time $t$  is defined by 
$\bSigma_t^{-1} = \U_{t-1}' \B_t\U_{t-1} $ where: (i) 
$\U_{t-1}=\bSigma_t^{-1/2},$ the upper triangular, scaled Cholesky component of $\bSigma_{t-1}^{-1}$; (ii) 
the $q\times q$  $\B_t$ is matrix beta distributed independently of $\bSigma_{t-1},$ 
 with matrix beta parameters $(\beta d_{t-1}/2,(1-\beta)d_{t-1}/2)$ where $\beta\in (0,1]$ is a chosen discount factor and $d_{t-1}=n_{t-1}+q-1.$
    The evolution implies 
$(\bSigma_t|\cD_{t-1})  \sim \tIW(n_t^*,\D_t^*)$ where $n_t^*=\beta n_{t-1}-(1-\beta)(q-1)$ and $\D_t^*=\beta\D_{t-1}.$  The reduced d.o.f. 
represents discounted information (increased uncertainty) due to the evolution, while the 
harmonic mean $\D_t^*/(n_t^*+q-1) = \D_{t-1}/(n_{t-1}+q-1)$  is unchanged through the evolution.

\section{Matrix Normal-Inverse Wishart Summaries \label{app:allCNIW}} 

\subsection{Partitioned Inverse Wishart} 
Suppose the $q\times q$ s.p.d. matrix $\bSigma\sim\tIW(n,\D)$.

\paragraph{Partition and Transforms:} 
For positive integers $q_c,q_e$ with $q_c+q_e=q,$ conformably partition 
$$ \bSigma = \begin{pmatrix} \bSigma_c&\bSigma_{ce}\\ \bSigma_{ec}&\bSigma_e\end{pmatrix} \quad\textrm{and}\quad
\D = \begin{pmatrix} \D_c&\D_{ce}\\ \D_{ec}&\D_e\end{pmatrix} 
\quad\textrm{with}\quad \bSigma_{ce}=\bSigma_{ec}', \ \ \D_{ce}=\D_{ec}'.$$ 
The component matrices are: 
$q_c\times q_c$ diagonal blocks $\bSigma_c,\D_c$;  
$q_e\times q_e$ diagonal blocks $\bSigma_e,\D_e$;    
$q_e\times q_c$ lower left off-diagonal blocks $\bSigma_{ec},\D_{ec}.$ 
Define transformed matrices 
$$\bGamma_e = \bSigma_{ec}\bSigma_c^{-1} \quad\textrm{and}\quad \bPsi_e =  \bSigma_e - \bSigma_{ec}\bSigma_c^{-1}\bSigma_{ec}'.$$ 
\paragraph{Distributional summaries:} 
\begin{itemize} \itemsep-3pt 
\item $\bSigma_c\sim \tIW(n,\D_c)$ and  $\bSigma_e\sim \tIW(n,\D_e)$;
\item $\bGamma_e|\bPsi_e \sim  \tN(\D_{ec}\D_c^{-1},\bPsi_e,\D_c^{-1}),$ matrix normal;
\item $\bPsi_e \sim \tIW(n+q_c, \D_e - \D_{ec}\D_c^{-1}\D_{ec}').$  
\end{itemize} 

\subsection{(Matrix) Normal-Inverse Wishart\label{app:NIW}} 
The $p\times q$ matrix $\bTheta|\bSigma \sim \tN(\M,\C,\bSigma)$, matrix normal with conformable matrix mean $\M=\tE(\bTheta|\bSigma)\equiv \tE(\bTheta)$, 
column $p\times p$ variance matrix $\C$ and row $q\times q$ variance matrix $\bSigma.$   With $\bSigma\sim \tIW(n,\D)$ the joint distribution is
 $(\bTheta,\bSigma) \sim \tNIW(\M,\C,n,\D).$ 

 \paragraph{Partition and Transforms:} 
With $\bSigma$ partitioning and definitions in Appendix~\ref{app:IW},  conformably partition $\bTheta = (\bTheta_c,\bTheta_e)$ 
and $\M=(\M_c,\M_e)$, each with $p\times q_c$ and $p\times q_e$ blocks. 

\paragraph{Distributional summaries:} 
\begin{itemize} \itemsep-3pt 
\item $\bTheta_c|\bSigma \sim \tN(\M_c,\C,\bSigma_c)$ and $\bTheta_e|\bSigma\sim \tN(\M_e,\C,\bSigma_e)$;
\item $\bTheta_e|\bTheta_c,\bSigma \sim \tN( \M_e+(\bTheta_c-\M_c)\bGamma_e',\C,\bPsi_e)$. 
\end{itemize}

\subsection{A Class of Conditional Normal-Inverse Wishart (CNIW) Distributions\label{app:CNIW}} 
With definitions, notation and partitioning as in Sections~\ref{app:IW} and~\ref{app:NIW},
specify additional parameters as follows: 
 \begin{itemize} \itemsep-3pt 
 \item a $p\times q$ matrix $\Z = (\Z_c,\Z_e)$ where $\Z_c$ is $p\times q_c$  and $\Z_e$ is $p\times q_e$;
 \item a $p\times p$ s.p.d. matrix $\C_e$;
  \item  a d.o.f. $\cniwdf_e>0$; 
 \item a $q\times q$ s.p.d. matrix $\H$ whose partitioned form conformable with that of $\bSigma$ 
  has s.p.d. diagonal blocks $\H_c,\H_e$ and lower left off-diagonal block $\H_{ec}$.
\end{itemize} 
The resulting CNIW p.d.f. conditional on any value of $\bTheta_c$ is defined in the compositional form
\beq{app:CNIW} p(\bTheta_e,\bGamma_e, \bPsi_e| \bTheta_c )
   = p(\bTheta_e| \bTheta_c,\bGamma_e, \bPsi_e) p(\bGamma_e | \bPsi_e ) p(\bPsi_e)
\eeq
with conditional independencies implicit in the notation and  independently of $\bSigma_c.$ 
The terms here are:
\begin{itemize} \itemsep-3pt 
\item $\bTheta_e| \bTheta_c,\bGamma_e, \bPsi_e  \sim \tMN(\Z_e + (\bTheta_c - \Z_c)\bGamma_e', \C_e, \bPsi_e);$ 
\item  $\bGamma_e | \bPsi_e \sim \tMN(\H_{ec}\H_c^{-1}, \bPsi_e, \H_c^{-1})$, matrix normal; 
\item $ \bPsi_e \sim \tIW(\cniwdf_e, \H_e - \H_{ec}\H_c^{-1}\H_{ec}').$  
\end{itemize}
The notation is  $(\bTheta_e,\bGamma_e, \bPsi_e| \bTheta_c ) \sim  \tCNIW(\Z, \C_e, \cniwdf_e, \H | \bTheta_c).$

\paragraph{NIW-CNIW Distributions:}  Suppose $(\bTheta_c,\bSigma_c ) \sim\tNIW(\M_c,\C,\bSigma_c)$  with the $p\times q$ mean matrix 
$\M$ partitioned conformably with $\bSigma$, i.e., $\M = (\M_c,\M_e)$  where $\M_c$ is $p\times q_c$ and $\M_e$ is $p\times q_e.$ 

It follows that the  specific choices $\Z_c=\M_c$,  $\Z_e=\M_e$, $\C_e = \C$, $\H=\D$ and $ \cniwdf_e = n+q_c$  reduce the CNIW distribution to the implied distribution (conditional on $\bTheta_c$) under the NIW. That is, the implied joint distribution for $(\bTheta,\bSigma)$ is precisely $\tNIW(\M,\C,n,\D).$
 
More generally, other choices of the CNIW parameters provide more flexibility in representing information and uncertainty about the components of $(\bTheta,\bSigma)$  in the above partitioned forms; this allows a focus on potentially different information for the components.  In general, 
the resulting joint distribution over components is no longer NIW; it is referred to as a member of the class of NIW-CNIW distributions; the p.d.f. is 
$p(\bTheta_c,\bSigma_c) p(\bTheta_e,\bGamma_e, \bPsi_e| \bTheta_c )$ where the first term is NIW and the second the
 conditional p.d.f. of \eqn{app:CNIW}.

\subsection{Prior-to-Posterior Updating in MVDLMs with NIW-CNIW Priors \label{app:p2pNIW-CNIW} }
It turns out that the class of NIW-CNIW distributions defines conjugate priors for likelihood functions arising in MVDLMs as in of \eqn{mvdlm},
substantially extending the existing NIW theory.  The  theory is presented here in general, at one point in time and ignoring the time subscript for clarity in focusing on the basics of of prior-to-posterior updating. 

The observation equation is 
$\y' = \F' \bTheta + \bnu'$ with  $\bnu \sim \tN(\bzero, \bSigma).$ Partition  $\y' = (\y_c', \y_e')'$ conformably with the partitioned $\bSigma$ above, 
so that $\y_c$ is $q_c\times 1$ and $\y_e$ is $q_e\times 1.$  Then, standard conditional normal theory provides the p.d.f. $p(\y|\bTheta,\bSigma)$ in the compositional form 
\beq{p2pNIW-CNIW} 
p(\y|\bTheta,\bSigma) = p(\y_c|\bTheta_c,\bSigma_c) p(\y_e|\y_c,\bTheta_c,\bTheta_e,\bGamma_e,\bPsi_e)  
\eeq
with components
\begin{itemize}\itemsep-3pt
\item $(\y_c'|\bTheta_c,\bSigma_c) \sim \tN(\F'\bTheta_c,\bSigma_c)$, 
\item $(\y_e'|\y_c,\bTheta_c,\bTheta_e,\bGamma_e,\bPsi_e)  \sim \tN( \F'\bTheta_e + (\y_c'-\F'\bTheta_c)\bGamma_e', \bPsi_e).$ 
\end{itemize} 
Suppose now the prior for states and components of variance matrix parameters has the NIW-CNIW form. It turns out that the posterior conditional on $\y$ is also NIW-CNIW. 
 
\paragraph{Theorem 1.} Under the above model,  assume the defined NIW-CNIW prior
\begin{itemize}\itemsep-3pt
\item $(\bTheta_c,\bSigma_c ) \sim\tNIW(\M_c^*,\C^*,n_c^*,\D_c^*)$,
\item $(\bTheta_e,\bGamma_e, \bPsi_e| \bTheta_c ) \sim  \tCNIW(\Z^*, \C_e^*, \cniwdf_e^*, \H^* | \bTheta_c)$ independently of $\bSigma_c.$ 
\end{itemize} Here all prior parameters superscripted $*$ are known.   Then, the posterior is also NIW-CNIW, namely
\begin{itemize}\itemsep-3pt
\item $(\bTheta_c,\bSigma_c |\y ) \sim\tNIW(\M_c,\C_c,n_c,\D_c)$,
\item $(\bTheta_e,\bGamma_e, \bPsi_e| \y, \bTheta_c ) \sim  \tCNIW(\Z, \C_e,\cniwdf_e, \H | \bTheta_c)$ independently of $\bSigma_c.$ 
\end{itemize} 
The proof, with details of the evaluation of the posterior parameters, follows.  
 
\subsubsection*{Marginal NIW Model} The $c-$marginal model of \eqn{p2pNIW-CNIW} for data $\y_c$ alone, coupled with the marginal NIW prior on  $(\bTheta_c,\bPsi_c)$, defines a traditional matrix normal, inverse Wishart set-up. Hence the posterior is NIW with parameters $(\M_c,\C_c,n_c,\D_c)$  computed from the 
usual NIW updating equations, as follows.  

Define the $q_c-$vector point forecast $\f_c$ by $\f_c' =  \F'\M_c^*$,  the resulting forecast error vector 
$\e_c = \y_c - \f_c$, and the  $p-$vector of adaptive coefficients $\A_c = \C_c^*\F/v_c$ where $v_c = 1 + \F'\C_c^*\F$. 
The posterior NIW parameters are then 
$\M_c = \M_c^* + \A_c\e_c'$, $\C_c = \C_c^* -\A_t\A_t'v_c$, $n_c = n_c^*+1$ and $\D_c = \D_c^*+\e_c\e_c'/v_c.$ 
 
\subsubsection*{CNIW Model}   The $e|c-$conditional model of \eqn{p2pNIW-CNIW} for data $\y_e$ alone-- and assuming known values of
$(\y_c,\bTheta_c)$-- defines a conditional likelihood to update the specified CNIW prior for $\bTheta_e,\bGamma_e,\bPsi_e.$ 
The observation model p.d.f.  
$p(\y_e|\y_c,\bTheta_c,\bTheta_e,\bGamma_e,\bPsi_e)  $ defines the conditional likelihood function for $(\bTheta_e,\bGamma_e,\bPsi_e).$ 
 
Define the $q-$vector $\z$ via  $\z'= \y'-\F'\Z^*$ and the $p-$vector 
$\A_e = \C_e^*\F/v_e$ with $v_e = 1+\F'\C_e^*\F.$   
Then the posterior CNIW parameters are given by 
$\Z = \Z^*+\A_e\z'$, $\C_e = \C_e^* - \A_e\A_e'v_e$, $s_e=s_e^*+1$ and $\H = \H^* + \z\z'/v_e$.

\paragraph{Proof of CNIW Updating.} 

\def\wty{\widetilde{\y}}\def\wtf{\widetilde{\f}}\def\wtr{\widetilde{\r}}

Two steps identify (a) the conditional 
posterior for $\bTheta_e$ given $\bGamma_e,\bPsi_e$, and (b) the marginal posterior for $\bGamma_e,\bPsi_e.$
These are, of course, conditional on $\y,\bTheta_c$ and all prior parameters. 

\medskip\noindent{(a) \rm Conditional for $\bTheta_e$.} 

Define $\wty_e$ via $\wty_e' = \y_e' -(\y_c'-\F'\bTheta_c)\bGamma_e'$. The conditional likelihood function for $\bTheta_e$  has the form of a matrix normal arising from the dynamic regression 
$\wty_e'|\bTheta_c \sim \tN( \F'\bTheta_e,\bPsi_e)$ with \lq\lq synthetic'' outcome $\wty_e'$. The conditional prior
$\bTheta_e \sim \tMN(\Z_e^* + (\bTheta_c - \Z_c^*)\bGamma_e', \C_e^*, \bPsi_e)$ is conjugate to this likelihood function hence the posterior is matrix normal. The parameter updates follow the general matrix normal theory used in MVDLMs, with results as follows. 

Define the synthetic forecast vector $\wtf_e$ and corresponding forecast error 
$\wtr_e$ via $\wtf_e' = \F'\{\Z_e^* + (\bTheta_c - \Z_c^*)\bGamma_e'\}$ and $\wtr_e=\wty_e-\wtf_e.$  
The standard theory shows that the conditional posterior for $\bTheta_e$ is 
\beq{proofa} \tN( \Z_e^* + (\bTheta_c - \Z_c^*)\bGamma_e'  + \A_e\wtr_e', \C_e, \bPsi_e)\eeq
where $\C_e = \C_e^* - \A_e\A_e'v_e$ with $\A_e = \C_e^*\F/v_e$ and $v_e = 1+\F'\C_e^*\F.$   

Now, 
$$\wtr_e' = \{\y_e' -(\y_c'-\F'\bTheta_c)\bGamma_e'\}  - \F'\{\Z_e^* + (\bTheta_c - \Z_c^*)\bGamma_e'\} = 
 \z_e'- \z_c'\bGamma_e'$$
where $\z_e'=  \y_e'-\F'\Z_e^*$ and $\z_c' =  \y_c'-\F'\Z_c^*$.
 Importantly, the terms in $\bTheta_c$ cancel out in this expression for $\wtr_e'.$  Substituting for $\wtr_e'$ in 
the posterior mean in~\eqn{proofa} leads to the mean expressed as 
$$\Z_e^* + (\bTheta_c - \Z_c^*)\bGamma_e'  + \A_e(\z_e'- \z_c'\bGamma_e') = \Z_e+(\bTheta_c-\Z_c)\bGamma_e'$$
where $\Z_e=\Z_e^*+\A_e\z_e'$ and $\Z_c=\Z_c^*+\A_e\z_c'$. 
Hence the posterior mean has the CNIW form  $\Z_e + (\bTheta_c - \Z_c)\bGamma_e'$ where  $\Z = (\Z_c,\Z_e)$ is
given by $\Z = \Z^*+\A_e\z'$ with $\z'=(\z_c',\z_e') = \y'-\F'\Z^*$.

\medskip\noindent{(a) \rm Marginal for $\bGamma_e,\bPsi_e$.} 

Consider now the  distribution of $\y_e$  marginalized over the conditional matrix normal for  $\bTheta_e$.  
This is normal with mean vector for $\y_e'$ given by 
$$\F'\{\Z_e^*+(\bTheta_c-\Z_c^*)\bGamma_e' \} + (\y_c'-\F'\bTheta_c)\bGamma_e' = 
    \F'\Z_e^* + \z_c'\bGamma_e'$$ 
with $\z_c'=\y_c'-\F'\Z_c^*$ as earlier defined.  Importantly, this mean vector for $\y_e'$  does not now involve $\bTheta_c$. The corresponding 
 variance matrix is $v_e\bPsi_e$ with scalar $v_e =1+\F'\C_e^*\F$ also earlier defined.   As all quantities here are known apart from 
  $(\bGamma_e,\bPsi_e),$ this normal p.d.f. results in a likelihood function for these parameters that has a matrix normal, inverse Wishart form.  This likelihood function is   easily seen to reduce to
  $$|\bPsi_e|^{-1/2} \etr\{-\bPsi_e^{-1}(\bGamma_e\z_c\z_c'\bGamma_e'-\bGamma_e\z_c\z_e'-\z_e\z_c'\bGamma_e'+\z_e\z_e' )/(2v_e) \}$$
   with $\z_e'=\y_e'-\F'\Z_e^*$ also as earlier defined. The NIW prior
  has p.d.f. proportional to 
  $$|\bPsi_e|^{-(q_e+s_e^* /2)} \etr\{-\bPsi_e^{-1}(\bGamma_e\H_c^*\bGamma_e'-\bGamma_e\H_{ec}^{*'}-\H_{ec}^*\bGamma_e'+\H_e^*)/2\}$$
  hence the product of the likelihood and prior has the NIW form with updated d.o.f.  
  $s_e=s_e^*+1,$ and  updated $q\times q$ s.p.d. matrix $\H$ whose diagonal blocks are $\H_c=\H_c^*+\z_c\z_c'/v_e$ and  
  $\H_e=\H_e^*+\z_e\z_e'/v_e$ and with  
   lower left off-diagonal block $\H_{ec}=\H_{ec}^*+\z_e\z_c'/v_e$.     This can be simply written as $\H = \H^* + \z\z'/v_e$ 
   where $\z'=(\z_c',\z_e')$ as earlier defined. 
  
   This completes the proof and delivers simple updating formul\ae\ in the CNIW setting.

\end{appendices}

\subsection*{Acknowledgements} 

The research reported here was developed while Graham Tierney was a PhD student in Statistical Science at Duke University,  while Christoph Hellmayr was Research Scientist at  $84.51^\circ$, and with  partial financial support from $84.51^\circ$.   All opinions, findings, conclusions and recommendations expressed in this paper are wholly those of the authors and do not reflect the views of $84.51^\circ$. 
 

\small\setlength{\bibsep}{4pt}
\bibliographystyle{chicago}
\bibliography{KevinLiEtAlcausalMVDLMs2024}

\begin{thebibliography}{}

\bibitem[\protect\citeauthoryear{Antonelli and Beck}{Antonelli and
  Beck}{2023}]{antonelli2023heterogeneous}
Antonelli, J. and B.~Beck (2023).
\newblock Heterogeneous causal effects of neighbourhood policing in {N}ew
  {Y}ork {C}ity with staggered adoption of the policy.
\newblock {\em Journal of the Royal Statistical Society, Series A\/}.

\bibitem[\protect\citeauthoryear{Ba{\'{n}}bura, Giannone, and
  Reichlin}{Ba{\'{n}}bura et~al.}{2010}]{BanburaEtAl2010}
Ba{\'{n}}bura, M., D.~Giannone, and L.~Reichlin (2010).
\newblock Large {B}ayesian vector autoregressions.
\newblock {\em Journal of Applied Econometrics\/}~{\em 25}, 71--92.

\bibitem[\protect\citeauthoryear{Brodersen, Gallusser, Koehler, Remy, and
  Scott}{Brodersen et~al.}{2015}]{brodersen2015inferring}
Brodersen, K.~H., F.~Gallusser, J.~Koehler, N.~Remy, and S.~L. Scott (2015).
\newblock Inferring causal impact using {B}ayesian structural time-series
  models.
\newblock {\em The Annals of Applied Statistics\/}~{\em 9\/}(1), 247--274.

\bibitem[\protect\citeauthoryear{Corradi and Guagnano}{Corradi and
  Guagnano}{1993}]{Corradi1993}
Corradi, F. and G.~Guagnano (1993).
\newblock Missing data and forecasting in multivariate time series: {An}
  application of the common components dynamic linear model.
\newblock {\em Journal of the Italian Statistical Society\/}~{\em 2}, 193--211.

\bibitem[\protect\citeauthoryear{Gruber and West}{Gruber and
  West}{2016}]{GruberWest2016}
Gruber, L.~F. and M.~West (2016).
\newblock {GPU}-accelerated {B}ayesian learning and forecasting in simultaneous
  graphical dynamic linear models.
\newblock {\em Bayesian Analysis\/}~{\em 11}, 125--149.

\bibitem[\protect\citeauthoryear{Gruber and West}{Gruber and
  West}{2017}]{GruberWest2017}
Gruber, L.~F. and M.~West (2017).
\newblock Bayesian forecasting and scalable multivariate volatility analysis
  using simultaneous graphical dynamic linear models.
\newblock {\em Econometrics and Statistics\/}~{\em 3}, 3--22.
\newblock arXiv:1606.08291.

\bibitem[\protect\citeauthoryear{Koop and Korobilis}{Koop and
  Korobilis}{2010}]{KoopKorobilis2010}
Koop, G. and D.~Korobilis (2010).
\newblock Bayesian multivariate time series methods for empirical
  macroeconomics.
\newblock {\em Foundations and Trends in Econometrics\/}~{\em 3}, 267--358.

\bibitem[\protect\citeauthoryear{Koop and Korobilis}{Koop and
  Korobilis}{2013}]{KoopKorobilis2013}
Koop, G. and D.~Korobilis (2013).
\newblock Large time-varying parameter {VARs}.
\newblock {\em Journal of Econometrics\/}~{\em 177}, 185--198.

\bibitem[\protect\citeauthoryear{Menchetti and Bojinov}{Menchetti and
  Bojinov}{2022}]{menchetti2022estimating}
Menchetti, F. and I.~Bojinov (2022).
\newblock Estimating the effectiveness of permanent price reductions for
  competing products using multivariate {B}ayesian structural time series
  models.
\newblock {\em The Annals of Applied Statistics\/}~{\em 16}, 414--435.

\bibitem[\protect\citeauthoryear{Pang, Liu, and Xu}{Pang
  et~al.}{2022}]{pang2022bayesian}
Pang, X., L.~Liu, and Y.~Xu (2022).
\newblock A {B}ayesian alternative to synthetic control for comparative case
  studies.
\newblock {\em Political Analysis\/}~{\em 30}, 269--288.

\bibitem[\protect\citeauthoryear{Papadogeorgou, Menchetti, Choirat, Wasfy,
  Zigler, and Mealli}{Papadogeorgou et~al.}{2023}]{papadogeorgou2023evaluating}
Papadogeorgou, G., F.~Menchetti, C.~Choirat, J.~H. Wasfy, C.~M. Zigler, and
  F.~Mealli (2023).
\newblock Evaluating {F}ederal policies using {B}ayesian time series models:
  {E}stimating the causal impact of the hospital readmissions reduction
  program.
\newblock {\em Health Services and Outcomes Research Methodology\/}~{\em 1},
  1--19.

\bibitem[\protect\citeauthoryear{Prado, Ferreira, and West}{Prado
  et~al.}{2021}]{PradoFerreiraWest2021}
Prado, R., M.~A.~R. Ferreira, and M.~West (2021).
\newblock {\em Time Series: Modeling, Computation \& Inference\/} (2nd ed.).
\newblock Chapman \& Hall/CRC Press.

\bibitem[\protect\citeauthoryear{Quintana and West}{Quintana and
  West}{1987}]{Quintana1987}
Quintana, J.~M. and M.~West (1987).
\newblock An analysis of international exchange rates using multivariate
  {DLMs}.
\newblock {\em The Statistician (now: Journal of the Royal Statistical Society,
  Series D)\/}~{\em 36}, 275--281.

\bibitem[\protect\citeauthoryear{Tierney, Hellmayr, Li, Barkimer, and
  West}{Tierney et~al.}{2024a}]{TierneyEtAl2024}
Tierney, G., C.~Hellmayr, K.~Li, G.~Barkimer, and M.~West (2024a).
\newblock Multivariate {B}ayesian dynamic modeling for causal prediction.
\newblock {\em Submitted for publication\/}.
\newblock arXiv:2302.03200.

\bibitem[\protect\citeauthoryear{Tierney, Hellmayr, Li, Barkimer, and
  West}{Tierney et~al.}{2024b}]{TierneyEtAl2024Supplement}
Tierney, G., C.~Hellmayr, K.~Li, G.~Barkimer, and M.~West (2024b).
\newblock Multivariate {B}ayesian dynamic modeling for causal prediction:
  {M}ore on models, data and analyses.
\newblock {\em Duke University: Statistical Science Technical Report\/}.
\newblock Available online at: {www.stat.duke.edu/$\sim$mw/mwrefs.html}.

\bibitem[\protect\citeauthoryear{Uhlig}{Uhlig}{1994}]{uhlig94}
Uhlig, H. (1994).
\newblock On singular {W}ishart and singular multivariate beta distributions.
\newblock {\em Annals of Statistics\/}~{\em 22}, 395--405.

\bibitem[\protect\citeauthoryear{West}{West}{2020}]{West2020Akaike}
West, M. (2020).
\newblock Bayesian forecasting of multivariate time series: {S}calability,
  structure uncertainty and decisions (with discussion).
\newblock {\em Annals of the Institute of Statistical Mathematics\/}~{\em 72},
  1--44.

\bibitem[\protect\citeauthoryear{West and Harrison}{West and
  Harrison}{1986}]{West1986}
West, M. and P.~J. Harrison (1986).
\newblock Monitoring and adaptation in {B}ayesian forecasting models.
\newblock {\em Journal of the American Statistical Association\/}~{\em 81},
  741--750.

\bibitem[\protect\citeauthoryear{West and Harrison}{West and
  Harrison}{1989}]{West1989}
West, M. and P.~J. Harrison (1989).
\newblock Subjective intervention in formal models.
\newblock {\em Journal of Forecasting\/}~{\em 8}, 33--53.

\bibitem[\protect\citeauthoryear{West and Harrison}{West and
  Harrison}{1997}]{west:harri:97}
West, M. and P.~J. Harrison (1997).
\newblock {\em Bayesian {F}orecasting and {D}ynamic {M}odels\/} (2nd ed.).
\newblock Springer.

\end{thebibliography}

\end{document}